\documentclass[12pt]{article}

\def\one{{\hbox{1\kern-.8mm l}}}

\def\be{\begin{equation}}
\def\ee{\end{equation}}
\def\bea{\begin{eqnarray}}
\def\eea{\end{eqnarray}}
\newcommand{\ket}[1]{|{#1}\rangle}

\newcommand{\braket}[2]{\langle{#1}|{#2}\rangle}
\makeatletter \def\section{\@startsection {section}{1}{\z@}{-3.5ex
plus -1ex minus -.2ex}{2.3ex plus .2ex}{\large\bf}}
\def\subsection{\@startsection{subsection}{2}{\z@}{-3.25ex plus -1ex
minus -.2ex}{1.5ex plus .2ex}{\normalsize\bf}}

\begin{document}
\begin{titlepage}
\rightline{NORDITA-2001/94 HE}  \rightline{DSF-41/2001}
\vskip 1.2cm
\centerline{\Large \bf $\mathbf{{\cal N}=2}$ Four-Dimensional Gauge Theories}
\vskip .4cm
\centerline{\Large \bf From Fractional Branes}

\vskip 1.4cm

\centerline{\bf M. Bertolini$^a$, P. Di Vecchia$^a$ and R. Marotta$^b$}

\vskip .8cm
\centerline{\sl $^a$ NORDITA, Blegdamsvej 17, 2100 Copenhagen \O, Denmark} 
\centerline{\tt teobert@nbi.dk, divecchi@alf.nbi.dk}
\vskip .5cm
\centerline{\sl $^b$ Dipartimento di Scienze Fisiche, Universit{\`{a}} di Napoli}
\centerline{\sl Complesso Universitario Monte S. Angelo, Via Cintia. I-80126 Napoli, Italy}
\centerline{\tt Raffaele.Marotta@na.infn.it}

\vskip 1.2cm

\begin{abstract}
This is a pedagogical and extended version of the results
published in Refs.~\cite{Bertolini:2000dk,d3d7} and presented by the
authors in various talks during the last year. We discuss the type II
D-branes (both regular and fractional) of the orbifold $R^{1,5}
\otimes R^4/Z_2$, we determine their corresponding supergravity
solution  and show how this can be used to study the
properties of  ${\cal{N}}=2$ super Yang-Mills. Supergravity is 
able to  reproduce the perturbative moduli space of
the gauge theory, while it does not encode the non-perturbative
corrections. The short distance region of space-time, which
corresponds to the infrared region of the gauge theory, is excised by an
enhan\c{c}on mechanism, and more states  should be included in the low
energy effective action in order to enter inside the enhan\c{c}on 
and recover the instanton corrections.
\end{abstract}
\end{titlepage}

\newpage

\tableofcontents

\renewcommand{\thefootnote}{\arabic{footnote}}
\setcounter{footnote}{0}
\setcounter{page}{1}
\section{Introduction}
\label{1}
Since the observation made by 't Hooft~\cite{THOOFT} of studying QCD
by using the large $N$ expansion it has been a dream of many particle
physicists to use it for studying with analytical methods the
non-perturbative properties of QCD as for instance confinement and
chiral symmetry breaking. Moreover, since the large $N$ expansion is
an expansion in the topology of the diagrams as string theory, it has
been a long standing hope to get a string theory for hadrons coming
out in the non-perturbative analysis of QCD. This has crashed,
however, with the fact that all known string theories contain gravity,
while QCD is a theory in flat Minkowski space.   

The Maldacena~\cite{mal} conjecture provides for the first time a
strong evidence that a string theory comes out from a gauge theory. It
states that four-dimensional ${\cal{N}}=4$ super Yang-Mills in flat
space is  equivalent to type IIB string theory compactified on $AdS_5
\times S^5$ and, since the two theories  live in two completely
different spaces, one does not run in the contradiction mentioned
above. On the other hand one expects that the emergence of a string
theory be related to confinement, while ${\cal{N}}=4$ super Yang-Mills
is a conformal invariant theory in a Coulomb phase and therefore does
not confine. Nevertheless, by means of the Maldacena conjecture one
has been able to obtain non trivial
informations~\cite{zare1,zare2,zare3,zare4}  on the strong coupling
behaviour of  ${\cal{N}}=4$ super Yang-Mills. 

In the last few years many attempts have been made to use  brane
dynamics for studying more realistic gauge theories. In particular,
the next in order of difficulty, namely ${\cal{N}}=2$ super Yang-Mills
(analyzed also at a non-perturbative level by Seiberg and
Witten~\cite{SW}), has been studied in terms of classical solutions of
the supergravity equations of motion corresponding to wrapped
branes~\cite{Bertolini:2000dk,d3d7,Polchinski:2000mx,marco,dario,zaffa}
of various  type~\footnote{For other  earlier  approaches see Refs.
\cite{Petrini,Pol}.}.

One of these approaches~\cite{Bertolini:2000dk,Polchinski:2000mx} is
based on using fractional D3-branes 
\cite{Douglas:1997xg,doug2,Diaconescu:1998br,marco1} of the orbifold $R^{1,5} 
\otimes R^4 /Z_2$. In this case the corresponding complete classical solution
of the equations of motion of type IIB supergravity has been obtained
and has been used in a probe analysis for deriving the moduli space of
${\cal{N}}=2$ super Yang-Mills that is known from Ref.~\cite{SW}.
Although the classical solution has a naked singularity at short
distances, it turns out that this does not cause any problem because
there is a distance, known as the enhan{\c{c}}on \cite{enhanc}, which
is bigger than the one where the singularity arises, where brane
probes become tensionless and where, correspondingly, the classical
supergravity solution looses meaning and the singularity is excised.
From the point of view of the gauge theory living on the world-volume
of  fractional D3-branes, the enhan\c{c}on corresponds to the scale
where the  gauge coupling constant diverges (the analogue of
$\Lambda_{\mbox{\tiny QCD}}$  in QCD). This means that using
supergravity  one can indeed reproduce the perturbative region of the moduli
space, obtaining,  for instance, the correct $\beta$-function. 
But, since at the enhan\c{c}on the classical solution becomes inconsistent, 
it is not possible to go further and use it for getting the non-perturbative 
instanton corrections of the Seiberg-Witten moduli space.

Another approach~\cite{dario,zaffa} is based on D5-branes wrapped on
supersymmetric two-cycles of non compact four dimensional Calabi-Yau
manifold, as ALE spaces~\footnote{A different approach based on 
wrapped branes is discussed 
in Refs.~\cite{ansar,Brinne:2000fh}}. The classical solution is obtained by lifting
to ten dimensions a solution found in $7$-dimensional gauged
supergravity. Although this approach is meant to give directly the
near-horizon limit of the brane, providing the supergravity dual \`a
la Maldacena, it turns out that it, as the one based on fractional
branes, is again only able to reproduce the perturbative behaviour of
the gauge theory living on the brane, since the enhan\c{c}on locus is
present also in these cases.

The previous results, that seem to be in strong contrast 
with a duality interpretation \`{a} la Maldacena where the supergravity 
solution gives a good description of the gauge theory for {\it large} 't 
Hooft coupling, can instead be easily understood if we regard the 
classical supergravity solution as an effective way of summing over 
all open string loops, as explained in detail in Ref.~\cite{nbps}.
From this point of view, in fact, one does not take the near-horizon limit 
({\it i.e.} $r \rightarrow 0$, where $r$ is the distance from the 
source branes) that anyway cannot be taken because of the
enhan{\c{c}}on, but rather expands the classical solution
around $ r \rightarrow \infty$ where the metric is almost flat and the 
supergravity approximation is valid. This expansion corresponds
to summing closed string diagrams at tree level, but, because of 
the open/closed string duality, it is also equivalent to summing
over open string loops. Therefore, expanding the supergravity
solution around $r\to \infty$ is equivalent to perform an expansion
for {\it small} 't Hooft coupling.

In view of these considerations, it is then not surprising that the 
supergravity solutions corresponding to fractional and wrapped branes
encode the perturbative properties of the ${\cal{N}}=2$
gauge theory living on their world-volume, 
and at the same time it is also natural that this approach
does not include the non-perturbative instanton corrections to the 
moduli space. The open and fascinating problem is then how to obtain
them from the brane dynamics. 

This is a pedagogical and extended version of the results published in
Refs.~\cite{Bertolini:2000dk,d3d7} in collaboration with Marialuisa
Frau, Alberto Lerda and Igor Pesando. We have written it for
commemorating Michael Marinov. One of us (PdV) has met him few times
in Soviet Union during the Meetings organized by Nordita that allowed
the physicists from the Soviet Union and those from the Western
Countries to meet and discuss in an extremely friendly and relaxed
atmosphere in the time of the cold war where many people as Michael
suffered of its consequences. After his migration to Israel he visited
Nordita a couple of times expressing  his happiness for his new life
there, but also his sadness for missing the life in Moscow.

The paper is organized as follows. In sections \ref{2} and \ref{3} we
discuss in great detail, respectively, the spectrum of massless closed
string states and that of massless open string states having their
endpoints on fractional and bulk  D3-branes of  the orbifold $R^{1,5}
\otimes R^4 /Z_2$. Section \ref{4} is devoted to the construction of
the boundary state describing fractional D$p$-branes and to its use to
compute their boundary action and the large distance behaviour of the
corresponding classical solution. In section \ref{5} we show that
fractional branes can be thought of as wrapped branes on vanishing
exceptional two-cycles of the corresponding orbifold. Sections \ref{6}
and \ref{7} are devoted to study the constraints imposed by supersymmetry on
the classical supergravity solution corresponding to the fractional 
D3-branes, and to the derivation of the solution itself. Finally, in the last
section, by probing the supergravity background that we have obtained, with
suitable fractional D-brane probes, we derive the properties of the
gauge theory living on bulk and fractional branes.

\section{Massless closed string states in orbifold 
$\mathbf{R^{1,5}\otimes R^4/Z_2}$}
\label{2}
Let us consider type II string theory on the orbifold $R^{1,5} \otimes
R^4/Z_2$ where $Z_2$ acts on the four directions  $x^6, x^7, x^8, x^9$
by changing their sign:
\begin{equation}
\{ x^6, x^7, x^8, x^9 \} \rightarrow \{ -x^6, -x^7, -x^8, -x^9 \}
\label{z276}
\end{equation} 
In this section we study the spectrum of the closed string states of
both type IIA and IIB theories.

We analyze the spectrum of closed strings in the light-cone gauge
where the classification group is $SO(8)$ that is obtained from the
original $SO(1,9)$ by dropping the string coordinates $x^0$ and
$x^1$. In the case of the orbifold $R^4/Z_2$ this group  is broken to:
\begin{equation}
SO(8) \rightarrow SO(4) \times SO(4)_{INT}
\label{bro89}
\end{equation}
where the orbifold group $Z_2$ acts on $SO(4)_{INT} $. Let us remember
that in an orbifold we have both untwisted and twisted sectors. The
former corresponds to the identity of the orbifold group, and consists
of the subset  of the states already present in flat space, which are
even under the orbifold group. The number of twisted sectors, instead,
depends on the orbifold under  consideration and is equal to the
number of non-trivial elements of the  discrete orbifold group. In our
case, where the orbifold group is $Z_2$, there is only one twisted
sector.

Let us start looking at the spectrum of the NS-NS sector, that is the
same for both type IIA and type IIB. The massless states of this
sector are given by:
\begin{equation}
\psi^{M}_{-1/2} {\tilde{\psi}}^{N}_{-1/2} |0,k \rangle
\label{nsns76}
\end{equation}
where $M$ and $N$ are indices of $SO(8)$ taking the values $M,N=2,3
\dots 9$. According to the breaking in eq.(\ref{bro89}) we write  $M =
( a,m)$ and $N= (b,n)$ where $a,b = 2,3,4,5$ are indices of the
space-time $SO(4)$ while $m,n =6,7,8,9$ are indices of $SO(4)_{INT}$.

Since the orbifold acts on the fermionic coordinate $\psi$ in the same
way as on the bosonic ones, according to eq.(\ref{z276}), in order to
preserve world-sheet supersymmetry, it is easy to see that the only
states that are even under $Z_2$ and that therefore survive the
orbifold projection are the following:
\begin{equation}
\psi^{a}_{-1/2} {\tilde{\psi}}^{b}_{-1/2} |0,k \rangle~~~~,~~~~
\psi^{m}_{-1/2} {\tilde{\psi}}^{n}_{-1/2} |0,k \rangle
\label{even65}
\end{equation}
Since both  $ \psi^{a}_{-1/2}$ and $ {\tilde{\psi}}^{b}_{-1/2}$
transform as the vector $(2,2)$ representation of $SO(4)$ and as the
singlet $(1,1)$ of $SO(4)_{INT}$, while both  $ \psi^{m}_{-1/2}$ and $
{\tilde{\psi}}^{n}_{-1/2}$ transform as the singlet $(1,1)$ of $SO(4)$
and as the vector representation $(2,2)$ of $SO(4)_{INT}$, it is easy
to see that the first state in eq.(\ref{even65}) transforms as:
\[
( (2,2) , (1,1) ) \ \otimes ( (2,2) , (1,1) ) = ( (3+1, 3+1), (1,1))=
\]
\begin{equation} 
= (3,3) + (1,3) + (3,1) +(1,1) , (1,1)
\label{deco87}
\end{equation}
corresponding to a graviton represented by $(3,3)$, to a 2-form
potential represented by $(3,1) +(1,3)$ and to a dilaton represented
by the singlet $(1,1)$. All these fields are singlets of
$SO(4)_{INT}$. Since $SO(4) = SU(2)_L \times SU(2)_R$, in the previous
formul$\ae$ we have labelled a representation of $SO(4)$ with $(p,q)$
where $p$ [$q$] is the dimension of the representation of 
$SU(2)_L [ SU(2)_R]$.  Analogously, it can be seen that the second
state in eq.(\ref{even65}) contains only  $16$ scalars (singlet with
respect to the first $SO(4)$) that transform according to the
representations $(3,3) + (3,1) + (1,3) + (1,1)$ of $SO(4)_{INT}$. In
conclusion the untwisted NS-NS sector of both type IIA and IIB
theories contains a graviton, a dilaton, a two-form potential and $16$
scalars.
  
Let us consider now the untwisted R-R sector. In the light-cone gauge
we can limit ourselves to the Dirac matrices of  $SO(8)$ that satisfy
the Clifford algebra:
\begin{equation}
\{ \psi_{0}^{M}, \psi_{0}^{N} \} = \delta^{MN}~~~,~~~M,N =2,3 \dots 9
\label{cli76}
\end{equation}
It is convenient to introduce the raising and lowering operators:
\begin{equation}
d_{i}^{\pm} = \frac{1}{\sqrt{2}} \left[ \psi_{0}^{2i} \pm i
  \psi_{0}^{2i+1} \right]~~~,~~~ i= 1,2,3,4.
\label{ralo67}
\end{equation}
satisfying the algebra:
\begin{equation}
\{ d_{i}^{+} , d_{j}^{-} \} = \delta_{ij}
\label{comu9}
\end{equation}
For each $i$ we have two states denoted by $|{s_i} \rangle$ with  $s_i
= \pm \frac{1}{2}$ that are eigenstates of the number operator $N_i$:
\begin{equation}
N_i \equiv -i \psi_{0}^{2i} \psi_{0}^{2i+1} = d_{i}^{+} d_{i}^{-} -
\frac{1}{2}~~~,~~~N_i |{s_i} \rangle = s_i |{s_i}\rangle
\label{ni56}
\end{equation}
A spinor of $SO(8)$  can then be represented by the $16$ states:
\begin{equation}
|{s_1 , s_2 , s_3 , s_4}\rangle
\label{state67}
\end{equation}
The chirality operator $\Gamma$ of $SO(8)$ is given by the product of
all Gamma matrices:
\begin{equation}
\Gamma = 2^4 N_1 N_2 N_3 N_4~~~,~~~\Gamma^2 =1
\label{ga98}
\end{equation}
The $8$ states with chirality equal to $\Gamma = +1 (-1)$ are
characterized by the fact that:
\begin{equation}
\sum_{i=1}^{4} N_i = \mbox{even\,(odd)}
\label{chi87}
\end{equation}
It is important to notice that the space-time $SO(4)$ acts only on the
indices $i=1,2$, while $SO(4)_{INT}$ acts on the remaining indices
$i=3,4$.  If we now limit ourselves  only to one of the two $SO(4)$
and we use the convention where $(1,2)$ [$(2,1)$] corresponds to the
eigenvalue $(+1) [(-1)]$ of the ${\hat{\Gamma}}$ matrix of the group
$SO(4)$ (for instance ${\hat{\Gamma}} \equiv - \Gamma_2 \Gamma_3 \Gamma_4 \Gamma_5
= 4 N_1 N_2$ in the case of space-time $SO(4)$), it is
easy to see that its two spinor representations correspond to the
following states:
\begin{equation}
(1,2) \sim (\frac{1}{2}, \frac{1}{2}) +  (- \frac{1}{2}, -\frac{1}{2})
\label{12}
\end{equation}
and
\begin{equation}
(2,1) \sim (\frac{1}{2}, -\frac{1}{2}) +  (- \frac{1}{2}, \frac{1}{2})
\label{21}
\end{equation}
This implies that the $8$ states with chirality $+1$ and $-1$ are
given respectively by:
\begin{equation}
8_s = ((1,2), (1,2)) + ((2,1), (2,1))~~~,~~~8_c = ((1,2),(2,1)) +
((2,1),(1,2))
\label{sta45}
\end{equation}
The orbifold group $Z_2$ acts on the spinor in eq.(\ref{state67}) as
follows:
\begin{equation}
|{s_1 , s_2 , s_3 , s_4}\rangle  \rightarrow {\rm e}^{i \pi (s_3 +
s_4)}  |{s_1 , s_2 , s_3 , s_4}\rangle
\label{act65}
\end{equation}
This implies that $Z_2$ acts on the two spinors of $SO(4)_{INT}$ as
follows:
\begin{equation}
(1,2) \rightarrow - (1,2) ~~~,~~~ (2,1) \rightarrow (2,1)
\label{act47}
\end{equation}
and the spinors $8_s$ and $8_c$ are transformed under the orbifold
action as follows:
\begin{equation}
8_s \rightarrow - ((1,2), (1,2)) + ((2,1),(2,1))~~,~~ 8_c \rightarrow
((1,2), (2,1)) - ((2,1), (1,2))
\label{act36}
\end{equation}
We are now ready to study the spectrum of the untwisted R-R
sector. Let us start with the type IIA theory that contains two
spinors with opposite chirality. This means that we should consider
the product $8_s \times 8_c$, where $8_s$ and $8_c$ correspond
respectively to the left and right movers, and keep only the states
that are even under the orbifold group $Z_2$. In this case the states
that survive the orbifold projections are the following:
\[
((2,1),(2,1)) \times ((1,2), (2,1)) + ((1,2),(1,2)) \times
((2,1),(1,2)) =
\]
\begin{equation}
= ((2,2), (3+1,1)) + ((2,2), (1, 3+1))
\label{spe63}
\end{equation}
that correspond to $8$ vectors of the space-time $SO(4)$. In
conclusion, the untwisted R-R sector of type IIA contains $8$ vector
fields. Considering now type IIB, we should take the product of two
spinors with the same chirality. The states that are even under the
orbifold projections are:
\[
((2,1),(2,1)) \times ((2,1),(2,1)) +((1,2),(1,2)) \times ((1,2),(1,2))
=
\]
\begin{equation}
= ((3+1,1), (3+1,1)) + ((1,3+1),(1,3+1))
\label{spe45}
\end{equation}
corresponding to $4$ two-form potential and $8$ scalars.

The previous spectra for the untwisted R-R sectors can also be
obtained by restricting ourselves to the states appearing in type II
theories in flat space that are even under the orbifold
projection. For instance in type IIA theory in flat space we have two
R-R fields $C_M$ and $C_{MNP}$. The even ones under the orbifold
projection are $C_a$, $C_{abc}$ and $C_{amn}$ where according to the
notation explained at the beginning of this section $a, b,c$ are
indices outside of the orbifold and $m,n$ along the orbifold.  The
previous states correspond in the six-dimensional space outside the
orbifold to $7$ vectors and a $3$-form potential that in six
dimensions is dual to a vector. This means that we obtain $8$ vectors
as with the other method used above. The same procedure can also be
applied to the case of type IIB theory containing the R-R fields $C$,
$C_{MN}$ and $C_{MNPQ}$ with self-dual field strength. The states
surviving the orbifold projection are $C$, $C_{ab}$, $C_{mn}$,
$\frac{1}{2} C_{abmn}$ and $\frac{1}{2} C_{\ell m n r}$ where the
factor $1/2$ takes care of the self-dual field strength. Those are
precisely the states found with the previous method.

Before moving  to the twisted sectors let us consider the
supersymmetric charges that survive the orbifold projection. The
ordinary type IIA has  two supercharges that transform respectively as
$8_s$ and $8_c$:
\begin{equation}
Q \sim 8_s ~~~,{\widetilde{Q}} \sim 8_c
\label{sucha34}
\end{equation}
Because of their transformation properties  under $Z_2$ (see
eq. (\ref{act36})) the states that are even under the orbifold
projection are the following:
\begin{equation}
Q \sim ((2,1),(2,1))~~,~~{\tilde{Q}} \sim ((1,2),(2,1))
\label{susyre}
\end{equation}
This shows that the orbifold $R^4 /Z_2$ keeps only $\frac{1}{2}$ of
the supersymmetry of flat space.  Proceeding in the same way in the
case of type IIB theory we get that the supersymmetric charges
surviving the orbifold projection are:
\begin{equation}
Q, {\widetilde{Q}} \sim ((2,1),(2,1))
\label{sucha45}
\end{equation}
It is interesting to notice that the supercharges in eqs.(\ref{susyre}) 
and (\ref{sucha45}) that survive the orbifold
projection all transform according to the representation $(2,1)$ of
$SO(4)_{INT}$ that  is left invariant under the action of the following
operator:
\begin{equation}
\Gamma^6 \Gamma^7 \Gamma^8 \Gamma^9 = - 4 N_3 N_4~~~,~~~\Gamma^{M} \equiv
\sqrt{2} \psi_{0}^{M}
\label{gso57}
\end{equation}
This follows trivially from the definition of the state $(2,1)$ given
in  eq.(\ref{21}).
 
Let us consider now the twisted sectors, starting from NS-NS
fields. In this case the massless states are spinors of $SO(4)_{INT}$
that we require to be even under both the action of the orbifold group 
and that of the  GSO operators:

\begin{equation}
P^L_{GSO} = P^R_{GSO} = \frac{1 + \Gamma^6 \Gamma^7 \Gamma^8
\Gamma^9}{2}
\label{gso78}
\end{equation}
where $L$ and $R$ label respectively the left and right movers of the
closed string.  As a consequence one gets:
\begin{equation}
((1,1),(2,1)) \times ((1,1),(2,1)) = ((1,1),(3+1,1))
\label{twispe34}
\end{equation}
corresponding to $4$ scalars. One is a singlet of $SO(4)_{INT}$, while
the other three are a triplet with respect to one of the two $SU(2)$
groups of $SO(4)_{INT}$.

Considering now the massless states in the R-R twisted sector, in type IIA 
theory we must consider the following GSO operators:
\begin{equation}
P^L_{GSO} = \frac{1 + \Gamma^2 \Gamma^3 \Gamma^3 \Gamma^4}{2} \quad , \quad 
P^R_{GSO} = \frac{1 - \Gamma^2 \Gamma^3 \Gamma^3 \Gamma^4}{2}
\label{gso56}
\end{equation}
that, together with the orfibold projection, select the following tensor product:
\begin{equation}
((2,1), (1,1)) \times ((1,2),(1,1)) = ((2,2),(1,1))
\label{vec56}
\end{equation}
that corresponds to a vector field. In the case of type IIB, instead,
we have the following tensor product:
\begin{equation}
((2,1),(1,1)) \times ((2,1),(1,1)) = ((3+1,1),(1,1))
\label{iib34}
\end{equation}
corresponding to a scalar and a self-dual two-form potential.
 
The orbifold we are considering has a curvature singularity at its
fixed point, corresponding to $x^6 = x^7 = x^8 = x^9 =0$. It is well
known that this singularity can be interpreted in terms of a vanishing
two-cycle ${\cal C}_1$ of a smooth ALE manifold. The twisted fields
can then  be understood as arising from the $p$-form fields appearing
in type II theories, dimensionally reduced on this vanishing two-cycle
(notice that  since the volume of the cycle is zero, there are no
Kaluza-Klein states, other than  the zero modes). In the NS-NS sector,
the antisymmetric two-form $B_{MN}$ gives rise to the scalar of the
NS-NS twisted sector that is a singlet of $SO(4)_{INT}$, see
eq.(\ref{twispe34}). The other three scalars, transforming  as a
triplet of $SU(2)$, are instead geometric moduli, related to the
metric tensor. In the R-R sector we get in type IIA theory a vector
field corresponding to dimensional reduction of the three-form
potential ($C_3 = {\cal A}_1  \wedge \omega_2$, where $\omega_2$ is
the differential form dual to the vanishing two-cycle ${\cal C}_1$)
and in type IIB a scalar field and a two-form potential with self-dual
field strength corresponding respectively to the dimensional reduction
of $C_2 = c \,\,\omega_2$  and $C_4 = {\cal A}_2 \wedge \omega_2$. It
is probably worth noticing that when we lift type IIA theory to
M-theory, the NS-NS scalar singlet provides an extra component to the
R-R twisted vector obtaining a vector in the $7$-dimensional space
orthogonal to the orbifold.

\section{Massless open string states in orbifold 
$\mathbf{R^{1,5} \otimes R^4/Z_2}$}
\label{3}

In this section we determine the spectrum of open strings ending on
the  D3-branes of the orbifold $R^{1,5} \otimes R^4 /Z_2$. The group
$Z_2$ consists of two generators: $g$  acting on the coordinates of
$R^4$ as in eq.(\ref{z276}), and its square that is the identity $e$.
If we consider a D3-brane located at a generic point of the orbifold
covering space, we must also include its image and consequently we
have  four kinds of open strings. Two kinds corresponding to open
strings having both their end-points on the brane or on its image and
two other kinds corresponding to open strings having one endpoint on
the brane and the other on its image and vice-versa. These four kinds
of open strings are described by a two by two Chan-Paton matrix that
we denote by:
\begin{equation}
\lambda = \left(
\begin{array}{cc}  
 \mbox{D3-D3}  & \mbox{D3-D3'} \\ \mbox{D3'-D3} & \mbox{D3' -D3'}
\end{array} 
\right)
\label{ma45}
\end{equation}
where each entry describes one of the four kinds of open strings. A
generic open string state in the NS sector will then be described by
the product of a  Chan-Paton matrix that we denote by $\lambda$ and an
oscillator state with a certain momentum along the world-volume of the
D3-brane. In particular, a massless state of the NS sector will have
the following form:
\begin{equation}
\lambda \,\psi_{-1/2}^{M} | 0, k \rangle~~,~~M= 0,1 \dots 9
\label{male45}
\end{equation}
The open string states that are allowed in an orbifold are those that
are left invariant under the action of $Z_2 $ that acts on both the
oscillators and the Chan-Paton factors. Since, in order to keep
world-volume supersymmetry, $Z_2 $ acts on the fermionic coordinates
in the same way as on the bosonic ones, the oscillator part of the
state in eq.(\ref{male45})  transforms under $g$  as follows:
\begin{equation}
\psi_{-1/2}^{\alpha,i} | 0, k \rangle \rightarrow
\psi_{-1/2}^{\alpha,i} | 0, k \rangle~~;~~~\alpha =0,1,2,3~~;~~i= 4, 5
\label{tra34}
\end{equation}
and
\begin{equation}
\psi_{-1/2}^{m} | 0, k \rangle \rightarrow - \psi_{-1/2}^{m} | 0, k
\rangle~,~m=6, \dots 9
\label{tra35}
\end{equation}
where we have denoted with $\alpha$ the world-volume directions of the
D3-brane, with $m$ the four directions along the orbifold and with $i$
the transverse ones outside the orbifold. On the other hand, the
Chan-Paton factors transform as follows under $Z_2 $:
\begin{equation}
\lambda \rightarrow \gamma(h) \lambda \gamma(h)^{-1}~~,~~ \gamma (e) =
\one = \left( \begin{array}{cc} 1 & 0\\ 0 & 1
\end{array}
\right) ~~,~~ \gamma (g) = \sigma_1 = \left( \begin{array}{cc} 0 & 1\\
1 & 0
\end{array}
\right)
\label{tra37}
\end{equation}
The matrix $\gamma (g)$ can be determined by requiring that it
exchanges an open string ending on the D3-brane with an open string
ending on its image and vice-versa. It is easy to check that the matrix
$\sigma_1$ in eq.(\ref{tra37}) satisfies this property.

Taking into account the action of the orbifold group on both the
oscillators and the Chan-Paton factors, one gets the following
invariant states that survive the orbifold projection:
\begin{equation}
\frac{1+ \sigma_1}{2} \otimes \psi_{-1/2}^{\alpha,i} |0, k\rangle
~~,~~ \frac{1- \sigma_1}{2} \otimes \psi_{-1/2}^{\alpha,i} |0, k
\rangle
\label{gau54}
\end{equation}
corresponding to two gauge fields living on the world-volume of the
D3-brane represented by the index $\alpha$ and four real Higgs fields
represented by the index $i$ and:
\begin{equation}
\frac{ \sigma_3 + i \sigma_2}{2} \otimes \psi_{-1/2}^{m} |0, k\rangle
~~~,~~~ \frac{\sigma_3 - i \sigma_2}{2} \otimes \psi_{-1/2}^{m} |0, k
\rangle~~~m=6,7,8,9
\label{gau57}
\end{equation}
corresponding to $8$ scalars. At the orbifold fixed point all these
fields are massless and are grouped together in two ${\cal{N}}=2$
vector multiplets, containing a gauge and two real Higgs fields each,
and two hypermultiplets, containing $4$ scalars each.

The action of $Z_2 $ on the Chan-Paton factors given in
eq.(\ref{tra37}) corresponds to the regular representation of $Z_2$
defined by the relation:
\begin{equation}
[ R (h)]_{h_1 h_2} = \delta_{h h_1, h_2}
\label{regu87}
\end{equation}
It turns out that it is a reducible representation as any non
one-dimensional representation of a discrete abelian group, as $Z_2$
is. In order to see this directly, it is convenient to perform a
change of basis in the space of the Chan-Paton factors $\lambda$ by
means of  the following transformation:
\begin{equation}
\lambda \rightarrow A^{-1} \lambda A~~,~~ A = \frac{1- i
\sigma_2}{\sqrt{2}}
\label{tra39}
\end{equation}
In this new basis the regular representation becomes:
\begin{equation}
\gamma (e) = \one~~~,~~~ \gamma (g) = \sigma_3
\label{basis34}
\end{equation}
and the massless states are given by:
\begin{equation}
A_{1}^{\alpha,i} \equiv \frac{1 + \sigma_3}{2} \otimes
\psi^{\alpha,i}_{-1/2} |0,k \rangle~,~ A_{2}^{\alpha,i} \equiv \frac{1
- \sigma_3}{2} \otimes \psi^{\alpha,i}_{-1/2} |0,k \rangle
\label{ve45}
\end{equation}
corresponding to the two gauge fields and the four Higgs scalar of the
two vector multiplets and by:
\begin{equation}
\Phi_{1}^{m} \equiv \frac{\sigma_1 + i \sigma_2}{2} \otimes
\psi_{-1/2}^{m} |0,k \rangle ~~,~~\Phi_{2}^{m} \equiv \frac{\sigma_1 -
i \sigma_2}{2} \otimes \psi_{-1/2}^{m}|0,k \rangle
\label{hyp59}
\end{equation}
corresponding to the two hypermultiplets. They can be grouped together
in the $2 \times 2$ matrix:
\begin{equation}
\left( \begin{array}{cc} A_1  & \Phi_1 \\ \Phi_2 & A_2
\end{array} \right)  
\label{ma34}
\end{equation}
The charge of the scalar hypermultiplets can be determined by the
commutation relations between the Chan-Paton factors of the gauge
vectors and those of the scalar fields. From the commutators:
\begin{equation}
[ \frac{1+ \sigma_3}{2} , \frac{\sigma_1 \pm i \sigma_2}{2}] =   \pm
\frac{\sigma_1 \pm i \sigma_2}{2}~~,~~ [ \frac{1- \sigma_3}{2} ,
\frac{\sigma_1 \pm i \sigma_2}{2}] =   \mp \frac{\sigma_1 \pm i
\sigma_2}{2}
\label{commu76}
\end{equation}
one gets that $\Phi_1 $ has charges $(1, -1)$ and $\Phi_2$ has
opposite charges $(-1, 1)$ with respect the two gauge fields $A_1$ and
$A_2$. Summarizing, the low energy effective theory living on
$N$ D3-branes is  four-dimensional ${\cal{N}}=2$ super Yang-Mills with gauge
group $U(N) \otimes U(N)$ and with two hypermultiplets transforming in
the bifundamental  representation of the two gauge groups:
\begin{equation}
\Phi_1 \sim (N, {\bar{N}})~~,~~\Phi_2 \sim ({\bar{N}}, N )
\label{tra43}
\end{equation} 
Such a theory is conformal invariant as it can be easily checked,
since the two $\beta$-functions are indeed vanishing. Hence the gauge
theory living on a D3-brane transforming according to the regular
representation  of the orbifold group, is conformal invariant. Notice
that the hypermultiplets  scalars  are associated with the possibility
of moving the D3-brane in the  orbifold directions, while the vector
multiplet scalars are associated to displacements along the fixed
plane ($x^4,x^5$). Bulk branes on orbifolds are then not much different
from usual D-branes in flat space. Indeed, when  moving a bound state
of $N$ bulk  D3-branes from the orbifold fixed point, only the
diagonal gauge group survives, and the corresponding low energy
effective  theory is equivalent to the Coulomb phase of $SU(N)$ ${\cal
N}=4$ super Yang-Mills, as it is the case for D3-branes in flat space.

In the new basis where the transformations of the $Z_2$ group are
given in eq.(\ref{basis34}), it is easy to see that the regular
representation is reducible, implying that the bulk branes transform
according to a reducible representation of the orbifold  group.  One
could look for more elementary branes transforming according to the
one-dimensional irreducible representations of the orbifold group. The group 
$Z_2$ has only two irreducible representations, ${\cal D}_I$ ($I=1,2$), 
given by:
\begin{equation}
\gamma_1 ( e ) = 1~~~,~~\gamma_1 (g) = 1
\label{irr45}
\end{equation}
and
\begin{equation}
\gamma_2 (e ) = 1~~~,~~~\gamma_2 (g) = -1
\label{irr43}
\end{equation}
The branes transforming according to one of the two previous
irreducible representations are called {\it fractional} branes. The
regular representation is of course the direct sum  of the above two
irreducible representations, namely:
\begin{equation}
R = \oplus \, {\cal D}_I \quad , \quad I=0,1
\label{reir2}
\end{equation}
This simple mathematical formula has in fact a very interesting
physical interpretation which will become clear when we will discuss
the closed  string interpretation of bulk and fractional branes in the
next section.

Since $Z_2$ has only two irreducible representations in this case 
there are only two kinds of fractional branes. Furthermore, being the
Chan-Paton factors one-dimensional, the fractional branes have the
property of living at the orbifold fixed plane $x^6 =x^7 = x^8= x^9
=0$, since  they do not have, by construction, an image.  Let us see
which is the low energy effective theory living on their
world-volume. In this case the massless open string states surviving
the orbifold projection are the following:
\begin{equation}
\psi^{\alpha}_{-1/2} |0,k \rangle~~~,~~\psi^{i}_{-1/2} |0,k \rangle
\label{ma43}
\end{equation}
corresponding in four dimensions to a gauge field and two real scalar
fields belonging to an ${\cal{N}} =2$ vector gauge multiplet. In the
case of a fractional brane the additional scalars belonging to the
hypermultiplets are projected out by the orbifold projection (this
implying that fractional branes are stuck on the orbifold fixed plane,
as already noticed). In conclusion the gauge theory living on $N$
fractional D3-branes of the orbifold $R^{1,5} \otimes R^4 / Z_2$ is
pure ${\cal{N}}=2$ super Yang-Mills with $U(N)$ gauge group, which  is
not conformal invariant. Therefore  fractional branes have the
advantage with respect to  bulk branes  that they allow for the study
of non-conformal gauge theories.

The previous analysis can be extended to any orbifold of the ADE
series \cite{Douglas:1996sw,Johnson:1997py}. For a generic orbifold of
the kind $R^4 / \Gamma$, ($\Gamma$ being a Kleinian subgroup of
$SU(2)$), bulk branes are defined as D-branes whose Chan-Paton factors
transform under the regular representation of $\Gamma$ (and hence, by
construction, they have images). Fractional branes, on the other hand,
are defined as D-branes whose  Chan-Paton  factors transform under the
irreducible representations of $\Gamma$ (and  do not have
images). Hence, for a generic orbifold theory, there are as many
different kinds  of fractional branes, as the number of different
irreducible representations of  $\Gamma$. While for abelian orbifolds
(A series, corresponding to $Z_N$)  the dimension of the irreducible
representations is one, for  non abelian  orbifolds (DE series), this
is not true anymore.  In these cases, the number  of different
fractional branes is then less than the order of $\Gamma$. The
generalization of eq.(\ref{reir2}) is indeed:
\begin{equation}
\label{reirn}
R = \oplus \, d_I {\cal D}_I \quad \mbox{with}  \quad \sum_{I=0}^{n-1}
d_I = |\Gamma| \quad , \quad I=0,2,\ldots,n-1
\end{equation}
where $|\Gamma|$ is the order of the discrete group $\Gamma$, $d_I$ is
the dimension of the $I$-th irreducible representation and $n$ is
their  number, this also being equal to the  number of different types
of fractional branes. Eqs. (\ref{reir2}) and (\ref{reirn})  seem to
suggest that bulk branes can somehow be thought as the 'sum' of
fractional branes. This n\"aive idea turns out to be correct, as it
will become apparent  in the next section, when discussing fractional
branes from the boundary state point of view. One can generalize the
analysis of the massless open string spectrum performed for $Z_2$ to
the case of a general orbifold group $\Gamma$ and easily see that the
gauge theory leaving on $N$ bulk branes corresponds to the following
group:
\begin{eqnarray}
U(d_0 N)\times U(d_1 N) \times \, ... \,\times U(d_{n-1} N)
\end{eqnarray}
with hypermultiplets transforming in the bifundamental of any given
couple  of gauge groups. Also in these more general cases, as for the
$Z_2$ orbifold,  it can be shown that the gauge theory living on  bulk
branes is conformal invariant, namely that all the $n$
$\beta$-functions are vanishing. The hypermultiplets correspond,
again,  to open string stretched between a D-brane and its images and
therefore are  present only in the low energy spectrum of bulk
branes. On the contrary,  fractional branes, which do not have images,
are described as before by pure  ${\cal N}=2$ and are stuck at the
orbifold fixed plane. They are free to move only on the fixed plane
($x^4,x^5$), the corresponding degrees of freedom  being  described by
the two scalars of the  ${\cal N}=2$ vector multiplet.

All previous considerations, which we have done for D3-branes, can  be
easily extended to a general bulk and fractional D$p$-brane. The only
essential difference, at this level, is that the low energy effective
theory living on them is in general a $p+1$ dimensional gauge
theory. Since we are mainly interested in four dimensional gauge
theories, we will not spend more time discussing D$p$-branes
here. Nevertheless, when discussing fractional branes from the closed
string point of view,  we will make a more general treatment which
will be valid for a generic value of $p$.

\section{Boundary state description of fractional branes}
\label{4}
In this section we  analyze in some detail the D$p$-branes of type II
string theories in the background of the orbifold $R^{1,5} \otimes R^4
/Z_2$ using the formalism of the boundary state.

The starting point in string theory for describing  a fractional
D$p$-brane is the vacuum energy $Z$ of the open strings stretched
between two fractional D$p$-branes that is given by:
\begin{equation}
Z = \int_0^\infty \frac{d s}{ s} ~{\rm Tr}_{\rm NS-R} \left[ \left(
\frac{1 + (-1)^F}{2}\right) \left(\frac{ e +g}{2} \right)  {\rm e}^{-2
\pi s (L_0-a)} \right]
\label{part45}
\end{equation}
where the first term under the trace performs the GSO projection, $e$
and $g$ are the two elements of   orbifold $Z_2$, and $a=1/2$ in the
NS sector and $a=0$ in the R sector.  When one takes the $e$ inside
the bracket, one gets half of the contribution of the open strings
stretched between two D$p$-branes  in flat space, whereas when one
takes the $g$ inside the bracket one  obtains the contribution of the
twisted sectors of the fractional D$p$-branes.  Let us consider in
general a D$p$-brane with $r+1$ directions of its world-volume outside
and $s =p-r$ directions along the orbifold $ R^4/Z_2$. To be more
specific, we divide both the world-volume and the transverse
directions in directions that are outside and along the orbifold. As
far as  the transverse directions are concerned we have then $4-s$
along the orbifold  and $5-r$ outside it. In this case the vacuum
amplitude is equal to:
\begin{equation}  
Z = Z_{e} + Z_{g}
\label{z}
\end{equation}
where:
\begin{eqnarray}
Z_{e} &=& \frac{1}{2} \int_0^\infty \frac{ds}{s}\, {\rm Tr}_{\rm
NS-R}\, \big[\, P_{\rm GSO}\, {\rm e}^{- 2 \pi s (L_0 -a)} \,\big]
\nonumber \\ &=& \frac{1}{2}\,
\frac{V_{p+1}}{\left(8\pi^2\alpha'\right)^{(p+1)/2}}\, \int_0^\infty
\frac{ds}{s^{(p+3)/2}}\, \frac{1}{2} \left[\, \frac{ f_3^8(q) -
f_4^8(q) - f_2^8(q)}{ f_1^8(q)}\, \right]
\label {z1} \\
Z_{g} &=& \frac{1}{2}\, \int_0^\infty \frac{ds}{s}\, {\rm Tr}_{\rm
NS-R}\,  \big[\,g\,  P_{\rm GSO} \,{\rm e}^{- 2 \pi s (L_0 -a)}\,
\big]  \nonumber \\ &=& \frac{V_{r+1}}{2^s \, \left(8 \pi^2
\alpha'\right)^ {(r+1)/2}}\, \int_0^\infty \frac{ds}{s^{(r+3)/2}}
\left[ \, \frac{f_3^4(q)\, f_4^4(q)}{f_1^4(q)\,f_2^4(q)} -
\frac{f_3^4(q)\,f_4^4(q)}{f_1^4(q)\,f_2^4(q)}\, \right]
\label{zg}
\end{eqnarray}
where $P_{\mbox{\tiny GSO}}$ is the GSO projection, $q= {\rm e}^{- \pi
s}$ and the $ f$'s are the standard one-loop  modular
functions. Notice the appearance of the important factor $2^{-s}$ in
eq.(\ref{zg}) that is due to the integration over the bosonic zero
modes along the orbifolded directions.

After performing the modular transformation  $s \to t=1/s $, $Z_e$ and
$Z_g$ can be interpreted  as tree level closed string amplitudes
between two untwisted and two twisted boundary states respectively,
that is:
\begin{eqnarray}
Z_{e} &=& \frac{\alpha' \pi}{2} \int_0^\infty dt \,\, \,^U\!\langle
{\rm D}p|\,{\rm e}^{- \pi t (L_0 + {\tilde{L}}_0 -2a )} |{\rm
D}p\rangle^U 
\label{z11} \\
 Z_{g} &=& \frac{\alpha' \pi}{2} \int_0^\infty dt \,\, \,^T\!\langle
{\rm D}p|\,{\rm e}^{- \pi t ( L_0 + {\tilde{L}}_0 ) } |{\rm
D}p\rangle^T 
\label{zc}
\end{eqnarray}
From eq.(\ref{z11}) it is immediate to realize that $Z_e$ is one half
of the  amplitude for D$p$-branes in flat space, and therefore the
untwisted part  of the boundary state is simply:
\begin{equation}
\label{bound1}  
\ket{{\rm D}p}^U =\,\, \frac{T_p}{2\sqrt{2}}\, \left(\, \ket{{\rm
D}p}_{\rm NS}^U \,+ \,\ket{{\rm D}p}_{\rm R}^U\, \right)~~,~~ T_p =
\sqrt{\pi} (2 \pi \sqrt{\alpha'})^{3-p}
\end{equation}
where $\ket{{\rm D}p}_{\rm NS}^U$ and $\ket{{\rm D}p}_{\rm R}^U$  are
the usual boundary states for a bulk D$p$-brane  given in
Refs.~\cite{bs,bs1}. Notice that in the previous equation we have
explicitly extracted from the boundary state of a bulk brane in flat
space the usual normalization factor $T_p/2$.
 
From eq.(\ref{zg}) we can see that the twisted amplitude for a
fractional D$p$-brane with $s$ directions along the orbifold is the
same as the one for a fractional D$r$-brane entirely outside the
orbifold, apart from a factor $2^{-s}$. Therefore, using
eq.(\ref{zc}), we can deduce that the boundary state $\ket{{\rm D}p}^T
$ is similar to the  boundary state for a fractional D$r$-brane
transverse to the orbifold, but with an extra factor of $2^{-s/2}$ in
its normalization. In conclusion, we get:
\begin{equation}
\label{boundt}  
\ket{{\rm D}p}^T \, = \,- \,\frac{1}{2^{s/2}}\,
\frac{T_r}{2\sqrt{2}\,\pi^2 \alpha' } \, \left(\,\ket{{\rm D}p}_{{\rm
NS}}^T + \ket{{\rm D}p}_{{\rm R} }^T\, \right)
\end{equation}
where:
\begin{equation}  
\label{proi}  
\ket{{\rm D}p}_{{\rm NS,R}}^T = \frac{1}{2}\,\left(\,\ket{{\rm
D}p,+}_{{\rm NS,R} }^T \,+\, \ket{{\rm D}p,-}_{{\rm NS,R} }^T\, \right)
\end{equation}
and the Ishibashi states  $\ket{{\rm D}p,\eta}_{\rm NS,R}^T$ are:
\begin{equation}
\label{bound2}  
\ket{{\rm D}p,\eta}^T_{\rm NS}=  \ket{{\rm D}p_X}^T\ket{{\rm
D}p_\psi,\eta}_{\rm NS}^T
\end{equation}  
in the NS-NS twisted sector, and:
\begin{equation}
\label{bound3}  
\ket{Dp, \eta}^T_{\rm R}=    \ket{Dp_X}^T\ket{Dp_\psi,\eta}_{\rm R}^T
\end{equation}
in the R-R twisted sector \footnote{In eqs. (\ref{bound2}) and
(\ref{bound3}) we omit the ghost and superghost contributions which
are not affected by the orbifold projection.}, with:
\begin{eqnarray}  
&&|{\rm D}p_X \rangle^T  = \delta^{(5-r)}({\widehat q}^i-y^i)
\prod_{n=1}^{\infty} {\rm e}^{-\frac{1}{n}   \alpha_{-n} \cdot S \cdot
\tilde\alpha_{-n}}    \,\prod_{r=\frac{1}{2}}^{\infty} {\rm e}^{-
\frac{1}{r} \alpha_{-r} \cdot S \cdot \tilde\alpha_{-r}}
\prod_{\alpha}{}^\prime\, | p_{\beta} =0 \rangle \prod_i{}^\prime\, |
p_i \rangle  \nonumber \\ \label{bound4}  \\
&&|{\rm D}p_{\psi} , \eta \rangle_{NS}^T  =
\prod_{r=\frac{1}{2}}^{\infty}  {\rm e}^{{\rm i}\eta\psi_{-r} \cdot S
\cdot    \tilde \psi_{-r} } \prod_{n=1}^{\infty} e^{{\rm i}\eta
\psi_{-n} \cdot S \cdot    \tilde \psi_{-n}} |{\rm D}p_{\psi} , \eta
\rangle ^{(0)\,\,T}_{\rm NS}   \\
\label{bound6}  
&&|{\rm D}p_{\psi} , \eta \rangle_{R}^T = \prod_{n=1}^{\infty} {\rm
e}^{{\rm i}\eta\psi_{-n} \cdot S \cdot     \tilde{\psi}_{-n} }
\prod_{r=\frac{1}{2}}^{\infty} {\rm e}^{{\rm i}\eta\psi_{-r} \cdot S
\cdot  \tilde{\psi}_{-r}} |{\rm D}p_{\psi} , \eta \rangle
^{(0)\,\,T}_{\rm R} 
\label{bou749}
\end{eqnarray}
where $S = (\eta_{\alpha \beta}, - \delta_{ij})$, with the
longitudinal indices $\alpha, \beta$ taking values $0,1, \ldots p$,
and the transverse indices $i,j$ taking values  $p+1, \ldots, 9$. The
prime in the vacuum of eq.(\ref{bound4})  indicates that the indices
$\beta$ and $i$ run only over the longitudinal and transverse
directions not included in the orbifold because there is no zero mode
on the orbifold directions.

The zero-mode part of the boundary state has a non trivial structure
in both sectors; in the NS-NS sector it is given by\cite{bs1}:
\begin{equation}
\label{bound7}  
|{\rm D}p_{\psi} , \eta \rangle ^{(0)\,\,T}_{\rm NS}=   \left(\widehat
C {\widehat{\gamma}}^{6} \dots {\widehat{\gamma}}^{5+s}  \frac{1 +
{\rm i}\eta\widehat\gamma}{1 + {\rm i}\eta}
\right)_{LM}|L\rangle|\widetilde M\rangle
\end{equation}
where $\widehat\gamma^\ell$ are the gamma matrices and $\widehat C$
the charge conjugation matrix of $SO(4)$,
$\widehat\gamma=\widehat\gamma^{6}...  \widehat\gamma^{9}$, and,
finally, $|L\rangle$ and $|\widetilde M\rangle$ are spinors of
$SO(4)$. The matrices of $SO (4)$  satisfy the following relations
under transposition:
\begin{equation}
{\widehat{C}}^{\,t} = {\widehat{C}}~~~~,~~~~
 {\widehat{\gamma}}^{\ell\,t} = {\widehat{C}}
 \,\,{\widehat{\gamma}}^{\ell} \,\, {\widehat{C}}^{-1}~~.
\label{transpo76}
\end{equation} 
In the R-R sector, instead, we have:
\begin{equation}
\label{bound8}  
|{\rm D}p_{\psi} , \eta \rangle ^{(0)\,\,T}_{\rm R}=   \left(\bar
C\bar\gamma^0...\bar\gamma^r\frac{1+ {\rm i}\eta\bar\gamma}   {1+{\rm
i}\eta}   \right)_{AB}|A\rangle|\widetilde B\rangle
\end{equation}
where $\bar\gamma^\alpha$ are the gamma matrices and $\bar C$ the
charge    conjugation matrix   of $SO(1,5)$,
$\bar\gamma=\bar\gamma^0...\bar\gamma^5$,   and, finally,  $|A\rangle$
and $|\widetilde B\rangle$ are spinors of $SO(1,5)$.  The matrices of
$SO(1,5)$ satisfy the following relations under transposition:
\begin{equation}
{\bar{C}}^t = - {\bar{C}}~~~~,~~~~ {{\bar{\gamma}}^{\alpha\,t}} = -
{\bar{C}}  \,\,{\bar{\gamma}}^{\alpha} \,\, {\bar{C}}^{-1}~~.
\label{transpo84}
\end{equation}
In order to compute the fermionic zero-mode contribution to $Z_g$ in
eq.(\ref{zc}) it is convenient to write explicitly the conjugate
vacuum states, which are given by\cite{bs1}:
\begin{equation}
{}^{(0) T}_{\rm NS} \langle {\rm D}p_{\psi} , \eta | = \langle
\widetilde M | \langle L | \left( \widehat C {\widehat{\gamma}}^{6}
\dots {\widehat{\gamma}}^{5+s}  \frac{1 - {\rm i}\eta\widehat\gamma}{1
- {\rm i}\eta}   \right)_{LM}
\label{bra36}
\end{equation}
for the twisted NS-NS sector, and:
\begin{equation}
{}^{(0) T}_{\rm R} \langle {\rm D}p_{\psi} , \eta | = \langle
\widetilde B | \langle A|  \left(\bar
C\bar\gamma^0...\bar\gamma^r\frac{1+ {\rm i}\eta\bar\gamma}   {1- {\rm
i}\eta}  \right)_{AB}
\label{brar34}
\end{equation}
for the twisted R-R sector. Using the previous expressions and
performing some straightforward  algebra, it is possible to show that:
\begin{equation}
{}^{(0) T}_{\rm NS} \langle {\rm D}p_{\psi} , \eta_1 |  {\rm
 D}p_{\psi} , \eta_2 \rangle ^{(0)\,\,T}_{\rm NS}  = 4 \delta_{\eta_1
 \eta_2;1}
\label{sandns39}
\end{equation}
for the NS-NS sector, and:
\begin{equation}
{}^{(0) T}_{\rm R} \langle {\rm D}p_{\psi} , \eta_1 |  {\rm D}p_{\psi}
 , \eta_2 \rangle ^{(0)\,\,T}_{\rm R} = - 4 \delta_{\eta_1 \eta_2;1}
\label{sandr28}
\end{equation}
for the R-R sector.

The previous twisted and untwisted boundary states are the building
blocks for constructing the boundary state associated with the two
kinds of fractional branes corresponding to the two irreducible
representations of the orbifold group $Z_2$ given in eqs.
(\ref{irr45}) and (\ref{irr43}) and that associated to a bulk
brane. Since, as one can see comparing eqs. (\ref{irr45}) and
(\ref{irr43}), the only difference between the two fractional
Dp-branes of $Z_2$ is the sign for the generator $\gamma (g)$, the
boundary states associated to them will just differ for a sign in
front of the twisted sector. This means that the boundary states
associated to the two fractional D-branes will be given by:
\begin{eqnarray}
\label{bosa}
\ket{{\rm D}p}_1 &=& \ket{{\rm D}p}^U + \ket{{\rm D}p}^T \\  \ket{{\rm
D}p}_2 &=& \ket{{\rm D}p}^U - \ket{{\rm D}p}^T 
\label{bos2}
\end{eqnarray}
On the other hand, a bulk brane is not coupled to the twisted sector
and the corresponding boundary state can be obtained by simply
summing-up  the boundary states of a fractional D$p$-brane of type 1
and one of type 2.  Indeed, by summing eqs.(\ref{bosa}) and
(\ref{bos2}) one sees that the twisted  contribution cancel and one is
left with 2 times the untwisted boundary state, which is precisely
that of a bulk brane: 
\begin{equation}
\ket{{\rm D}p}_{b} = \ket{{\rm D}p}_1 + \ket{{\rm D}p}_2 =  2 \,
\ket{{\rm D}p}^U 
\label{bos3}
\end{equation}
where the subscript ${\small b}$ in the last equation stands for {\em
bulk}. Since the tension of a brane is proportional to the
normalization of the corresponding boundary state, eq.(\ref{bos3})
shows that a fractional brane has a tension that is $1/2$ of that of a
bulk brane.  All previous considerations can be generalized to any
orbifold of the ADE series. In particular, eq.(\ref{bos3}) becomes in
these more general cases:
\begin{eqnarray}
\ket{{\rm D}p}_{b} &=& \sum_I\, \ket{{\rm D}p}_I
\label{bosn}
\end{eqnarray}
where the convention on the index $I$ is the same as in the previous
section.  Eqs.(\ref{bos3}) and (\ref{bosn}) are nothing else than the
closed string counterpart  of eqs.(\ref{reir2}) and
(\ref{reirn}). Once  again we see that in an orbifold theory the bulk
D$p$-branes can be thought  of as bound states of more elementary
D$p$-branes, the so-called fractional branes.

Having determined the boundary state for both the untwisted and the
twisted sectors of a fractional D$p$-brane, we will use it in the
following  for computing the couplings of the brane with  the closed
string fields. That will help us to determine the world-volume action
of a fractional D$p$-brane and the large distance behaviour of the
classical supergravity solution corresponding to it.  These are well
known things by now and the reader is urged to consult
Refs.~\cite{bs,nbps,antonella} for details and explanations.  In
particular, it is important to stress that by saturating the
previously constructed boundary states with the closed string states
one gets the couplings of the D-brane with the closed string states
that are canonically normalized in the bulk action written in the
orbifold covering space. We want, however, to write the couplings
corresponding to the fields defining in the physical space.  In
Refs.~\cite{Bertolini:2000dk,d3d7} we have normalized the bulk action
with an overall factor $1/2\kappa_{orb}^2$ where $\kappa_{orb} =
\sqrt{2}\, \kappa$ and we have taken $\omega_2$ in eq.(\ref{omega289})
to be normalized in such a way that the first integral in
eq.(\ref{omega289}) is equal to $\sqrt{2}$ instead of $1$ and the
second integral in the same equation is equal to $1$ instead of
$\frac{1}{2}$. In this paper we normalize the bulk action with an
overall factor $\frac{1}{2 \kappa^2}$ and we use an $\omega_2$
satisfying the relations in eq.(\ref{omega289}). 

By saturating the boundary state $|{\rm D}p\rangle$ with  the massless
closed string states of the various sectors, one can determine  which
are the fields that couple to the fractional D$p$-brane. In
particular, following the procedure found in Ref.~\cite{bs} and
reviewed in Ref.~\cite{antonella}, one can find that in the untwisted
sector the D$p$-brane emits the graviton $h_{\mu\nu}$\footnote{We
recall that the graviton  field and the metric are related by
$G_{\mu\nu}=\eta_{\mu\nu} + 2\kappa h_{\mu\nu}$ where $2 \kappa^2 =
(2\pi)^7 (\alpha')^4 g_{s}^{2}$. },  the dilaton $\phi$ and the
(p+1)-form potential $C_{p+1}$. For the case $s=0$, as for instance 
is the case for a fractional D3-brane, the couplings of these fields with
the boundary state are explicitly given by~\cite{anto}:   
\bea
\braket{{\rm D}p}{h} &=& -\frac{T_p}{2}\,\, h_{\alpha}^{\,\,\, \alpha}
\,V_{p+1}~~,\nonumber \\ \braket{{\rm D}p}{\phi} &=&
\,\frac{T_p}{2\kappa} \frac{3-p}{4}\,\, \phi\,V_{p+1}~~,\nonumber \\
\braket{{\rm D}p}{ C_{p+1}} &=&  \frac{T_p}{2 \kappa} \,C_{01\dots
p}\, V_{p+1}
\label{untw9}
\eea   
where $T_p =\sqrt{\pi} \,(2\pi\sqrt{\alpha '})^{(3-p)}$,
appearing in the normalization of the boundary state, is related to
the brane tension in units of the gravitational coupling constant
\cite{bs,bs1}, $V_{p+1}$ is the (infinite) world-volume  of the
D$p$-brane, and the index $\alpha$ labels its $p+1$ longitudinal
directions.
  
By doing this same analysis in the twisted sectors, we find that the
boundary state $|{\rm D}p\rangle$ emits a massless  scalar $\widetilde
b$  (the modulus related to $B_2$, namely the singlet under
$SO(4)_{INT}$  in eq.(\ref{twispe34})) from the NS-NS sector, and a
$(p+1)$-form potential $A_{p+1}$ from the R-R sector. These fields
exist only at the orbifold fixed point $x^6=x^7=x^8=x^9=0$,  and their
couplings with the boundary state turn out  to be given by~\cite{anto}:
\bea    
\braket{{\rm D}p}{\widetilde b} &=& \mp \frac{T_r}{2^{s/2}\,
2\,\kappa}\, \frac{1}{2\pi^2{\alpha'}} \,\widetilde b \,\, V_{r+1}
~~,\nonumber \\ \braket{{\rm D}p}{A_{p+1}} &=& \pm
\,\frac{T_r}{2^{s/2} \,2 \kappa}\, \frac{1}{2\pi^2{\alpha'}}\,A_{01
\dots p}\, V_{r+1}
\label{twi86}
\eea     
where $V_{r+1}$ is the (infinite) world-volume of the
D$p$-brane that lies outside the orbifold. The upper sign refers to
fractional branes  of type 1 while the lower sign to fractional branes
of type 2 (we recall, again, that on the orbifold we are considering
there are two types of fractional branes).

From the explicit couplings (\ref{untw9}) and (\ref{twi86}), it is
possible to infer the form of the world-volume action of a fractional
D$p$-brane. Of course, the boundary state approach allows to obtain
only the terms of the world-volume action that are linear in the bulk
fields.  However, terms of higher order can be determined with other
methods \cite{d3d7}. Numerically, both couplings  (\ref{untw9}) and
(\ref{twi86}) are the same as those we derived in  Refs.
\cite{Bertolini:2000dk,d3d7}, the only difference being, as  already
stressed, that here we are expressing them in terms of $\kappa$
instead of $\kappa_{orb}$. Therefore, we obtain (in the Einstein
frame):    
\be  
\left.S_{~\rm b}^{\rm Dp}\right|_{\rm U} =
-\,\frac{\tau_p}{2} \, \int d^{p+1}x~{\rm
e}^{(p-3)/4\,\phi}\,\sqrt{-\det G_{\alpha \beta}} ~+~
\frac{\tau_p}{2}\, \int C_{p+1}
\label{sbuntw}
\ee  
where $\tau_p \equiv T_p/{\kappa} =  \frac{ (2 \pi
\sqrt{\alpha'})^{-p}}{g_s \sqrt{\alpha'}} $  is the tension of the
bulk branes that is also equal to that of the branes in flat space and
$G_{\alpha\beta}$ is the induced metric.  It is easy to  check that
this action correctly accounts for the couplings (\ref{untw9}).

Eq. (\ref{sbuntw}) shows that fractional branes have a tension that in
the case of the orbifold under consideration is just a half of that of
a bulk brane. That is the reason of the name fractional
branes. Similarly we can see that its charge with respect to the R-R
field $C_{p+1}$ is a half of that carried by bulk D$p$-branes. As
already noticed,  the same conclusion holds from eq.(\ref{bos3}) and,
for more general orbifolds, from eq.(\ref{bosn}). Summarizing, for a
generic orbifold theory, calling $\tau_p$ and $\mu_p$ the tension and
the charge of a bulk D$p$-brane respectively, those of fractional
branes are:
\begin{eqnarray}
\label{tpf}
\tau_{p,I} &=& \frac{d_I}{|\Gamma|} \tau_p \,=\,  \frac{d_I}{|\Gamma|}
\frac{T_p}{\kappa} \\
\label{qpf}
\mu_{p,I} &=& \frac{d_I}{|\Gamma|} \mu_p \,=\, \frac{d_I}{|\Gamma|}
\frac{T_p}{\kappa} 
\end{eqnarray}
For the twisted fields, instead, things are slightly more
complicated. Using the couplings in eq.(\ref{twi86}) one can write 
\[
\left.S_{~\rm b}^{\rm Dp}\right|_{\rm T} = \pm \frac{1}{2 \pi^2
\alpha'}  \left\{ - \frac{{ \tau}_{r}}{2\,\cdot\, 2^{s/2}} \,\int
d^{r+1} x~ {\rm e}^{(p-3)/4\, \phi}  \sqrt{-\det
G_{\alpha\beta}}~{\widetilde b} \,+ \right.
\]
\be \left. \,+ \frac{{ \tau}_{r}}{2 \cdot \, 2^{s/2}}\,\int
{\cal{A}}_{r+1}+...\right\}
\label{sbtw}
\ee     
where in the first term the four-dimensional induced metric
has been inserted  to enforce reparametrization invariance on the
world-volume, while the ellipses stand for terms of higher order which
are not accounted by the boundary state approach but which, in
principle, can be present.

In the case of a fractional D$p$-brane with no world-volume directions
along the orbifold, we finally get the following boundary action for a
fractional brane of type 1~\cite{sabella}:
\[
S_1 = - \frac{\tau_p}{2} \int d^{p+1} x \,\, {\rm e}^{(p-3)/4\, \phi}
\sqrt{- \det G_{\alpha \beta}} \left( 1+ \frac{{\tilde{b}}}{2 \pi^2
\alpha'} \right) +
\]
\begin{equation}  
+ \frac{\tau_p}{2} \int_{V_{p+1}} \left[ C_{p+1} \left( 1+
\frac{{\tilde{b}}}{2 \pi^2 \alpha'} \right) +  \frac{1}{2 \pi^2
\alpha'} {\cal{A}}_{p+1}  \right]
\label{bou72}
\end{equation}
and the following action for a fractional brane of type 2:
\[
S_2 = - \frac{\tau_p}{2} \int d^{p+1} x \,\, {\rm e}^{(p-3)/4 \phi}
\sqrt{- \det G_{\alpha \beta}} \left( 1 - \frac{{\tilde{b}}}{2 \pi^2
\alpha'} \right) +
\]
\begin{equation}  
+ \frac{\tau_p}{2} \int_{V_{p+1}} \left[ C_{p+1} \left(1 -
\frac{{\tilde{b}}}{2 \pi^2 \alpha'} \right) -  \frac{1}{2 \pi^2
\alpha'} {\cal{A}}_{p+1}  \right]
\label{bou72b}
\end{equation}
The world-volume action of a bulk brane is given by the sum of the two
previous ones, namely $S_{b}=S_1 + S_2$, and is coincident with the
world-volume action of a D$p$-brane in flat space. In fact, the bulk
D$p$-branes of an orbifold are pretty much similar to the normal
branes in flat space and indeed are only coupled to closed string
fields of the untwisted sector as the metric, the dilaton and the R-R
field $C_{p+1}$. Fractional D$p$-branes are instead characterized
by the fact that they couple also to the fields of the twisted
sector. As already noticed, from the supergravity point of view twisted 
fields are the zero  modes of the usual NS-NS and R-R forms present in the 
type II spectrum  dimensionally reduced on the shrinking cycles of the
orbifold. Indeed, the orbifolds $R^4/\Gamma$ can be seen as singular
limit of smooth ALE spaces where the compact 2-cycles characterizing
these smooth spaces shrink to zero size. This can suggest some
relation between fractional branes and higher dimensional branes
wrapped on these exceptional cycles. We will come back on this issue
in the next section.

Let us end this  section by using the boundary state formalism to
compute the asymptotic behaviour of the various fields the fractional
branes couple to, in the corresponding classical brane solution (see
Ref.~\cite{bs} for an explanation of this technique). Considering a
stack of $N_1$ fractional D$p$-branes of the type 1  and
$N_2$ branes of type 2, we find  that, to leading order in $N_1g_s,
N_2g_s$, the metric is:
\begin{equation}
ds^2 \sim ( 1 - \frac{Q_p}{r^{7 -p}} \times \frac{7-p}{8} )
\eta_{\alpha \beta} dx^{\alpha} d x^{\beta}  +  ( 1 +\frac{Q_p}{r^{7
-p}} \times \frac{p+1}{8} ) \delta_{ij} dx^i dx^j
\label{me59}
\end{equation}
the dilaton is
\begin{equation}
\phi \sim \frac{3-p}{4} \frac{Q_p}{r^{7-p}}
\label{dil95}
\end{equation}
and the R-R untwisted field is:
\begin{equation}
C_{p+1} \sim - \frac{Q_p}{r^{7-p}} dx^0 \wedge dx^1 \dots \wedge dx^p
\label{rr76}
\end{equation}
where
\begin{equation}
Q_p \equiv \frac{ \kappa T_p}{(7-p) \Omega_{8-p}} \,(N_1 + N_2)~~,~~
\Omega_q = \frac{2 \pi^{(q+1)/2}}{\Gamma(\frac{p+1}{2})}~~,~~ r^2 =
\sum_i (x^i )^2
\label{qp876}
\end{equation}
The large distance behaviour of the twisted fields that are
stuck at the orbifold fixed point and therefore depend only on the
transverse directions outside of the orbifold, is given by:
\begin{equation}
{\tilde{b}} = \frac{K_p}{\rho^{3-r}}~~,~~K_p = \frac{2 \kappa
  T_r}{2^{s/2} (r-3)\, \Omega_{4-r}} \,\frac{1}{2\pi^2\alpha'}\, (N_1
  - N_2)
\label{bfi45}
\end{equation}
for the fluctuation of the $b$ field and
\begin{equation}
{\cal{A}}_{p+1} = - \frac{K_p}{\rho^{3-r}}  dx^0 \wedge dx^1 \dots
\wedge dx^p~~,~~ \rho^2 = x_{p+1}^{2} + \dots x_{5}^{2}
\label{rrtwi98}
\end{equation}
for the R-R twisted field. In section 7 we will write the equations of
motion of type IIB supergravity and restricting us to the case $p=3$ 
(and $r=0$) we will determine the complete supergravity solution 
describing a bound state of $N_1$ fractional D3-branes of type 1 and 
$N_2$ of type 2.

\section{Fractional branes as wrapped branes}
\label{5}
In this section we investigate at a deeper level the idea that we have
just anticipated, namely that fractional branes are just a certain
kind of wrapped branes. Let us first state this correspondence in a
precise mathematical fashion. Later, we will test its validity in more
concrete terms by comparing with the results obtained in the previous
section. The orbifolds $R^4/\Gamma$ are singular limits of ALE spaces,
the latter being non compact four dimensional manifolds uniquely
characterized, for any given $\Gamma$, by the presence of compact
holomorphic 2-cycles (which topologically are spheres) which shrink to
zero size in the orbifold limit. A well established mathematical
result (known as the McKay correspondence \cite{mk}) states that for
any given ALE space these 2-cycles are in one-to-one correspondence
with the simple roots $\alpha_I$  of a simply-laced Lie algebra (the
ADE {\it extended} Dynkyn diagrams) and these roots correspond to the
irreducible representations ${D\cal }_I$ of $\Gamma$. Actually, the
number of cycles equals the number of roots of the non-extended Dynkyn
diagrams, and hence is one less than the full number of roots and
irreducible representations. Indeed the trivial  irreducible
representation,  ${\cal D}_0$ (the one defined by eq.(\ref{irr45}),
for the $Z_2$ orbifold) is associated to a cycle ${\cal C}_0$ which is
minus the sum of all other cycles ${\cal C}_i$, ${\cal C}_0 =
-\sum_{i=1}^{n-1} d_i {\cal C}_i$. The corresponding simple root,
$\alpha_0$, is the extra root of the extended Dynkyn
diagram. Schematically one then has:
\begin{equation}
 \alpha_I \leftrightarrow {\cal C}_I \leftrightarrow {\cal D}_I \quad
\mbox{with} \quad I=0,1,\cdots,n-1
\end{equation}
Recalling from section 3 that fractional branes are uniquely
identified by the irreducible representations ${\cal D}_I$ of
$\Gamma$, one can then suspect the existence of a relation between
fractional branes and the shrinking cycles of the orbifold
$R^4/\Gamma$. This is indeed the case. The precise statement is as
follows. A fractional D$p$-brane is a D($p+2$)-brane wrapped on a
compact 2-cycle of a ALE manifold, in the limit in which the volume of
such cycle vanishes and the ALE space degenerates to the orbifold
$R^4/\Gamma$. These  branes can exist in the orbifold limit because,
although the size of the compact cycle shrinks to zero, a non
vanishing $B_2$-flux  persists on it, keeping the brane
tensionful. For a general orbifold the precise value of this flux
reads:
\begin{equation}
\label{bfluxn}
\int_{{\cal C}_i} B_2\,=\, (2\pi
\sqrt{\alpha'})^2\,\frac{d_i}{|\Gamma|}
\end{equation}
As already noticed, in the case of the $Z_2$ orbifold one has
$d_i/|\Gamma|=1/2$. This non-vanishing background flux  is not put by
hand but, as shown in \cite{bflux}, is required to keep string theory
conformal on the orbifold. It is this requirement which makes the
existence of fractional D-branes as stable non-perturbative states of
the string spectrum possible. As already noticed, there is one cycle
less than the number of irreducible representations. In fact, the
fractional D$p$-brane associated to the trivial representation ${\cal
D}_0$ is obtained by wrapping a D($p+2$)-brane on ${\cal C}_0$, with an
additional background flux of the world-volume gauge field ${\cal F}$
such that $\int_{{\cal C}_0} {{\cal F}}_2 = 2 \pi$. As we will
explicitly show in the case of $\Gamma=Z_2$, this assures that such a
brane gets an untwisted D$p$-brane charge of the same sign of that of
the branes associated to the non-trivial representations. This in fact
guarantees that it is a brane and not an anti-brane.

Let us now explicitly verify all these statements by considering our
working example, the orbifold $Z_2$. As already discussed, in this
case we have just one shrinking cycle, ${\cal C}_1$, and just two
different kinds of fractional D$p$-branes. The fractional brane of type
1 should correspond to a D($p+2$)-brane wrapped  on ${\cal C}_1$. The
fractional brane of type 2 to a D($p+2$)-brane wrapped on ${\cal C}_0 =
- {\cal C}_1$ and  with a non vanishing ${\cal F}$-flux on it. Let us
then consider the world-volume action of a wrapped D($p+2$)-brane and
see how it actually gives rise to the actions in eqs. (\ref{bou72})
and (\ref{bou72b}) in the limit of shrinking cycle. In the Einstein
frame a D($p+2$)-brane world-volume action has the form:
\begin{eqnarray}
S &=&  - \tau_{p+2} \int d^{p+3} x \,\,{\rm e}^{\frac{p-1}{4}\phi}
    \sqrt{- \det \left[ G_{\alpha\beta} + {\rm e}^{- \phi/2}
    \left(B_{\alpha\beta} + 2 \pi \alpha'  {\cal
    F}_{\alpha\beta}\right)\right]} + \nonumber \\ && +\tau_{p+2} \int
    \left( C \wedge {\rm e}^{B + 2 \pi \alpha' {\cal F}} \right)_{p+3}
\label{bp+2}
\end{eqnarray}
The smooth limit of the $Z_2$ orbifold is the well known Eguchi-Hanson
space, which  has an antiself-dual two-form $\omega_2$ which is
associated to the compact 2-sphere ${\cal C}_1$ whose radius we define
as $a$. We use conventions where  $\omega_2$ satisfies the following
properties:
\begin{equation}
\omega_2 = - {}^{*} \omega_2~~,~~ \int_{{\cal C}_1} \omega_2 = 1~~,~~
\int_{R^4 /Z_2} {}^* \omega_2 \wedge \omega_2 = \frac{1}{2}
\label{omega289}
\end{equation}
The compact cycle vanishes in the orbifold limit $a \rightarrow 0$
but, as already said,  a non-zero $B_2$-flux persists on it. In order
to obtain, from the action in eq.(\ref{bp+2}), the world-volume
actions of the two fractional D$p$-branes given in eqs.(\ref{bou72})
and(\ref{bou72b}) we should start from an action with  no world-volume
fields switched-on along the $p+1$ non-compact directions of the
world-volume. That is to say,  both $B$ and ${\cal F}$ are
non-vanishing only on the cycle ${\cal C}_1$. The action in
eq.(\ref{bp+2}) describes a brane wrapped on ${\cal C}_1$ by
considering the world-volume $V_{p+3}$ as a product of the
$p+1$-dimensional volume $V_{p+1}$ times the volume of the cycle
${\cal C}_1$ and keeping only those fields that are left in the limit
of $a \rightarrow 0$:
\begin{equation}
V_{p+3} = V_{p+1} \times {\cal C}_1~~,~~ B_2 = b
\,\omega_2~~,~~C_{p+3} = {\cal{A}}_{p+1} \wedge \omega_2
\label{wra98}
\end{equation}
By noticing that the metric has no support on the vanishing cycle, one
can easily factorize the matrix in the determinant in the action
(\ref{bp+2}) as a direct product of a  $(p+1) \times (p+1)$ matrix
$G_{\alpha\beta}$ times a $2\times 2$ matrix where only  $B$ and
${\cal F}$ are present. Let us consider first the case of a fractional
brane  of type 1. We want to show that it corresponds to a
D($p+2$)-brane wrapped on ${\cal C}_1$  with no ${\cal
F}$-flux. Inserting the expressions (\ref{wra98}) into eq.(\ref{bp+2})
one gets in this case:
\begin{eqnarray}
S &=&   \tau_{p+2} \Bigg\{ -\int d^{p+1} x \,\,{\rm
    e}^{\frac{p-3}{4}\phi} \sqrt{- \det \left( G_{\alpha\beta}\right)}
    \int_{{\cal C}_1} B +  \int C_{p+1}\int_{{\cal C}_1} B  + \int
    {\cal{A}}_{p+1} \Bigg\} \nonumber \\ &=&  \tau_{p} \Bigg\{ -\int
    d^{p+1} x \,\,{\rm e}^{\frac{p-3}{4}\phi} \sqrt{- \det \left(
    G_{\alpha\beta}\right)} \, \left(\frac{1}{2}+ \frac{1}{(2 \pi
    \sqrt{\alpha'})^2} \tilde{b} \right)  \nonumber \\ && +\, \int C_{p+1}
    \left(\frac{1}{2}+ \frac{1}{(2  \pi \sqrt{\alpha'})^2} \tilde{b}
    \right) + \frac{1}{(2 \pi  \sqrt{\alpha'})^2} \int {\cal{A}}_{p+1}
    \Bigg\}
\label{bpf1}
\end{eqnarray}
In the second step we have used the fact that:
\begin{equation}
(2 \pi \sqrt{\alpha'})^2\, \tau_{p+2} = \tau_{p} \quad \mbox{and}
\quad \int_{{\cal C}_1} B = b = (2 \pi \sqrt{
\alpha'})^2\,\left(\frac{1}{2} + \frac{1}{(2 \pi \sqrt{ \alpha'})^2}
\tilde{b} \right)
\label{b67}
\end{equation}
where $\tilde b$ is the fluctuation of the $B_2$-flux around the
background value given in eq.(\ref{bfluxn}). The above action
precisely coincides with that in eq.(\ref{bou72}), as anticipated. By
repeating the same reasoning for a D($p+2$)-brane which is  wrapped on
${\cal C}_0 = - {\cal C}_1$ but with an additional ${\cal F}$-flux
such that $\int_{{\cal C}_0} {{\cal F}}_2 = 2 \pi$ one easily gets:
\begin{eqnarray}
S &=&  \tau_{p+2} \Bigg\{-\int d^{p+1} x \,\,{\rm
    e}^{\frac{p-3}{4}\phi} \sqrt{- \det \left( G_{\alpha\beta}\right)}
    \int_{{\cal C}_0} \left( B + 2\pi\alpha' {\cal F}\right) \nonumber
    \\  &+& \int C_{p+1}\int_{{\cal C}_0}  \left( B + 2\pi\alpha'
    {\cal F}\right) + \int {\cal{A}}_{p+1} \Bigg\} \nonumber \\ &=&
    \tau_{p} \Bigg\{- \int d^{p+1} x \,\,{\rm e}^{\frac{p-3}{4}\phi}
    \sqrt{- \det \left( G_{\alpha\beta}\right)} \, \left(\frac{1}{2} -
    \frac{1}{(2 \pi \sqrt{\alpha'})^2} \tilde{b} \right)  \nonumber \\
    && +\, \int C_{p+1} \left(\frac{1}{2} - \frac{1}{(2 \pi
    \sqrt{\alpha'})^2} \tilde{b} \right) - \frac{1}{(2 \pi
    \sqrt{\alpha'})^2} \int {\cal{A}}_{p+1} \Bigg\}
\label{bpf2}
\end{eqnarray}
which is just the action in eq.(\ref{bou72b}). In deriving the above
equation it is  worth noting that:
\[
\int_{{\cal C}_0} \left(B \,+\, 2 \pi \alpha'  {\cal F}\right) = -
\int_{{\cal C}_1} B + 2 \pi \alpha' \int_{{\cal C}_0} {\cal F} =
\]
\begin{equation}
=- (2 \pi \sqrt{\alpha'})^2\,\left(\frac{1}{2}  + \frac{1}{(2 \pi
\sqrt{\alpha'})^2} \tilde{b} \right) + (2 \pi \sqrt{\alpha'})^2 = (2
\pi \sqrt{\alpha'})^2 \,\left(\frac{1}{2} - \frac{1}{(2 \pi
\sqrt{\alpha'})^2}  \tilde{b} \right)
\label{for968}
\end{equation}
From this last equation it is clear that, as anticipated, the presence
of the ${\cal F}$-flux has the effect of letting the asymptotic value
of the {\em untwisted} charge to be unchanged. By summing up the
two actions (\ref{bpf1}) and (\ref{bpf2}) one gets back the
world-volume  action of a bulk brane, according to the idea that bulk
branes, in an orbifold theory, can be thought of as bound states of
fractional branes of different kinds. Again, the procedure described
throughout this section for the the $Z_2$ orbifold, can be easily
extended to more  general $\Gamma$'s. We refer to \cite{marco} for a
complete treatment of these more general cases.

\section{Requirements of supersymmetry}
\label{6}
The goal of this section and the subsequent one is to obtain the
supergravity  solution describing (a bound state  of) fractional
D3-branes on the $Z_2$  orbifold. In this section we study the
constraints that supersymmetry  imposes on the solution. We do not do
this just for completeness, but because, as we shall see, supersymmetry
actually imposes certain conditions  on the fields entering the
solution which drastically simplify the  structure of the equations of
motion and, correspondingly, the derivation of the solution itself.

We are interested, as usual, in  classical supersymmetric backgrounds
where the dilatino $\lambda$ and the gravitino $\psi_{M} $ are
consistently put to zero. Moreover, in order to insure supersymmetry,
we require that the supersymmetry variations of both $\lambda$ and
$\psi_{M}$ be vanishing. In this way we will obtain some constraints
on the ansatz. A compact way of writing the gravitino and dilatino
variation is~\cite{susy}:
\[
\kappa \,\,\delta \psi_M  = \left( D_M - \frac{i}{2} Q_M  \right)
\epsilon + \frac{i}{16 \cdot 5!} F_{M_1\dots M_5}\Gamma^{M_1 ,   \dots
M_5} \Gamma_M \,\,\epsilon +
\]
\begin{equation} 
- \frac{1}{16} \left( 2 \not{\widetilde{G}}_{(3)} \Gamma_M +  \Gamma_M
\not {\widetilde{G}}_{(3)}\right) \epsilon^*  
\label{vargrav01} 
\end{equation}
and
\begin{equation}
\kappa \,\, \delta \lambda = i \left(  \not{P}  \epsilon^{*} -
\frac{1}{4} \not {\tilde{G}}_{(3)} \epsilon \right)   
\label{vardil01}
\end{equation}
where $\not{\tilde{G}}_{(3)}= \frac{1}{3!}\Gamma^{MNP}
\tilde{G}_{MNP}$, $D_M=\partial_M + \frac{1}{4}
\omega_{Mrs}\Gamma^r\Gamma^s$ is the covariant  derivative with
respect to the metric $g_{MN}$, 
\[
P_M= \frac{\partial_M B}{1-B\,B^*} \,\,\,\,\,\,\,\,\, Q_M=\frac{Im
(B\partial_MB^*)}{1-B\,B^*} 
\]
\begin{equation}
B= \frac{1+i\tau}{1-i\tau}  \,\,\,\,\,\,\,\,\, \tilde{G}_{(3)}=
i\left(\frac{1+i\tau^*}{1-i\tau}\right)^{1/2} e^{\phi/2}  G_3
\label{defsusy}
\end{equation}
and $\epsilon$ is a complex ten-dimensional spinor with definite
chirality: $\Gamma_{11}~\epsilon~=~-~\epsilon$, $\Gamma_{11} =
\Gamma^{\underline{0}} \dots \Gamma^{\underline{9}}$ and the
underlined indices are flat indices. Finally, the complex scalar
$\tau$ and the complex 3-form $G_3$ are: 
\begin{equation}
\tau = C_0 + i {\rm e}^{-\phi} \quad , \quad G_3 = F_3 + \tau H_3
\label{taug3}
\end{equation} 
where $F_3=dC_2$ and $H_3= dB_2$. Notice that fractional branes
actually couple to $G_3$ (or better to say, to  the twisted fields
arising from its dimensional reduction on the vanishing  cycle of the
orbifold, see previous sections), this being a specific feature of a
general class of supergravity solution recently discussed in the
literature (for an explicit example on smooth ALE spaces, see for 
instance Ref.\cite{Bertolini:2001ma}). For this reason, in the following 
we closely follow the approach discussed in Refs.~\cite{GRANA,GRANA2} 
where the supersymmetry constraints  for solutions with non trivial 
$G_3$-flux has been discussed. As far as our orbifold is concerned, 
these include both fractional D3 and D7 branes while D1 and D5 branes 
belong to a different  class of solutions (see the above cited references 
for details). An ansatz compatible with the symmetries of the system is:
\begin{equation}
ds^2 = Z^{-1/2} \eta_{\alpha\beta}dx^{\alpha}dx^{\beta} + Z^{1/2}
e^{-\phi(x^i)} \delta_{ij}dx^i dx^j + Z^{1/2}  \delta_{mn} dx^m dx^n  
\label{ansmetr} 
\end{equation}
and
\begin{equation}
{\tilde{F}}_5 = dH^{-1}\wedge V_4 + *\left( dH^{-1}\wedge V_4  \right)  
\label{ansf5}
\end{equation}
with $\alpha,\beta= 0\dots 3$, $i,j =4,5$, $m,n = 6\dots 9$. As far
as the transverse space is concerned, it is convenient to introduce
complex  coordinates $z^i$ ($i=1,2,3$) as:
\begin{equation}
z_1=x^4 + i x^5 \,\,\,\,\, z_2= x^6+ix^7\,\,\,\,\, z_3= x^8+ix^9
\label{compcoor}
\end{equation}
Let us start studying the dilatino variation. The vanishing of the
dilatino equation is obtained by a separate cancellation between the
two terms appearing in eq.(\ref{vardil01}):
\begin{equation}
\not{P}  \epsilon^{*}=0~~,~~ \not{\tilde{G}}_{(3)}\epsilon=0  
\label{vardil02}
\end{equation}
In the case of a D3-brane the first condition is simply solved by
considering a constant $\tau $. This is, however, not a good solution
for the D7-brane since it is coupled to $\tau$. On
the other hand, $\tau$ can depend only on the coordinates transverse
to the world-volume of the D7-brane, namely on $z_1$ and
${\bar{z}}_1$. But, if we assume that in this case $\tau $ is an
holomorphic function of  $z_1$ (namely  $\partial_{\bar 1} \tau =0$)
and impose the condition $\Gamma ^{1} \epsilon ^{*}=0$ (the index $1$
corresponds to the first complex variable introduced in
eq.(\ref{compcoor}) on the spinor $\epsilon$, it is easy to check
that the first condition in eq.(\ref{vardil02}) is again satisfied.

The second equation in eq.(\ref{vardil02}) fixes some components of $
{\tilde{G}}_{(3)}$ to be zero. In order to satisfy it we have to
extend the previous condition $\Gamma ^{1} \epsilon ^{*}=0$ to the
other two values of $i$. Moreover we assume that $ {\tilde{G}}_3$ has
only non-zero components if the indices are along the six-dimensional
space transverse to the D3-brane.  With these two assumptions the
second equation in (\ref{vardil02}) is satisfied if we impose\cite{GRANA2}:
\begin{equation}
\tilde{G}_{ijk}= \tilde{G}_{i\bar{i}k} =0~~,~~i,j,k =1,2,3  
\label{dilvar03}
\end{equation}
Let us now study the gravitino variation. By imposing, again, a
separate cancellation between the terms depending on
$\not{\tilde{G}}_{(3)}$ and the other ones in the
eq. (\ref{vargrav01}), we arrive at the following equations:
\begin{eqnarray}
&&\left( D_M - \frac{i}{2} Q_M\right) \epsilon + \frac{i}{ 5!\cdot 16}
F_{M_1\dots M_5}\Gamma^{M_1, \dots M_5}\Gamma_M \epsilon=0  
\label{vargrav04} \\
&& \left(2 \not{\tilde{G}}_{(3)} \Gamma_M  +  \Gamma_M \not
{\tilde{G}}_{(3)} \right) \epsilon^*=0  
\label{vargrav05}
\end{eqnarray}
Using eqs. (\ref{ansmetr}) and (\ref{ansf5}), eq.(\ref{vargrav04}),
once  specified for the longitudinal components of the D3-brane,
reduces to the  following one:
\begin{equation}
\partial_{\alpha} \epsilon -\frac{1}{8} \Gamma_w \Gamma_\alpha
\left[1- \frac{1}{2} \left(\frac{Z}{H} +\frac{H}{Z}\right)\Gamma_{(5)}
\right] \epsilon=0  
\label{vargrav06}
\end{equation}
where $\Gamma_w = \left( \gamma^i\partial_i+
\gamma^m\partial_m\right)\ln Z$,  $\Gamma_{(5)}= i
\Gamma^{\underline{0}} \dots \Gamma^{\underline{3}}$.
Eq.(\ref{vargrav06}) is clearly satisfied by choosing $Z=H$ and by
taking a spinor $\epsilon$ that does not depend on the coordinates of
the longitudinal  directions of the D3-brane and that has positive
four-dimensional chirality: $\Gamma_{(5)} \epsilon = \epsilon$. This
equation together with the condition $ \Gamma_{11} \epsilon = -
\epsilon$ implies that:
\begin{equation}
\Gamma_{(7)} \epsilon = - i \epsilon~~~,~~~\Gamma_{(7)} =
\Gamma^{\underline{4}} \dots \Gamma^{\underline{9}}
\label{six56}
\end{equation}  
This condition leads to the identity $\not{\tilde{G}}_{(3)}\epsilon
= i\not{\tilde{G}}_{(3)} \Gamma_{(7)}\epsilon $,  that, when solved, gives
the important relation \cite{GUB,cve3}:
\begin{equation}
- i\,\tilde{G}_{(3)} ={}^{*6} \tilde{G}_{(3)}  
\label{identity}
\end{equation}
where $*6$ denotes the Hodge dual in the six-dimensional transverse
space of a D3-brane.

In order to study the conditions that follow from the components
transverse to the D3-brane of eq.(\ref{vargrav04}) it is convenient to
decompose the $10$-dimensional Dirac matrices in terms of the
$4$ and $6$-dimensional ones by writing $\epsilon=
\xi\otimes \eta$, where $\xi$ and $ \eta$ are respectively spinors in
four and six dimensions satisfying the conditions:
\begin{equation}
\gamma^5 \xi = \xi~~~,~~~\gamma^7 \eta = - i \eta~~,~~\Gamma_{11}= -i
\gamma^5 \otimes \gamma^7
\label{con75}
\end{equation}
By using the previous decomposition in the transverse components in
eq.(\ref{vargrav04}) we obtain:
\begin{equation}
\partial_i \eta + \frac{1}{4} \partial_i \alpha(z_1) \,\eta -
\frac{1}{8} \partial_i \ln H \eta=0 
\label{cond52}
\end{equation}
and
\begin{equation}
\partial_{\bar{i}}\eta - \frac{1}{4} \partial_{\bar{i}}
\bar{\alpha}(\bar{z}_1) \,\eta - \frac{1}{8} \partial_{\bar i} \ln H \eta=0  
\label{finale}
\end{equation}
with $\alpha (z_1)+\bar{\alpha}({\bar{z}}_1)=\phi + \ln (1-BB^{*})$.
The previous system of equations is simply solved by choosing
$\eta=H^{1/8}e^{- (\alpha (z_1)-\bar{\alpha}(\bar{z}_1))/4}\chi$, being
$\chi$  a constant spinor.

Finally eq.(\ref{vargrav05}) fixes that some other components of
${\tilde{G}}_{(3)}$ are zero. In particular it imposes\cite{GRANA2}:
\begin{equation}
\tilde{G}_{\bar{i} \bar{j}\bar{k}}=\tilde{G}_{\bar{i} \bar{j} k}=0
\label{finale2}
\end{equation}
Collecting together eqs. (\ref{dilvar03}) and (\ref{finale2}) we
conclude that the only non-vanishing components of ${\tilde{G}}_{3}$
are  $\tilde{G}_{\bar{i}j k}$ with $i \neq j,k$. This implies that
$\tilde{G}_{(3)}$  is a (2,1) form. In the next section we will see 
that this property simplifies the equation of motions.

\section{Classical solution for fractional D-branes}
\label{7}
In this section, by considering type IIB supergravity on the $Z_2$
orbifold, we will derive the complete classical solution describing a bound
state of $N_1$ fractional branes of type 1 and $N_2$ of type 2. We will 
see that this solution belongs to a class of type IIB supersymmetric 
solutions all characterized by the presence of a non-trivial  $G_3$-flux.

Let us start by considering the action (in the Einstein frame) of type
IIB supergravity in ten dimensions that can be written
as~\footnote{Our conventions for curved indices and forms are the
following: $\varepsilon^{0\dots9}=+1$; signature $(-,+^9)$;
$\mu,\nu=0,\dots,9$; $\omega_{(n)}={1\over n!}  \,\omega_{\mu_1 \dots
\mu_n} dx^{\mu_1}\wedge\dots\wedge  dx^{\mu_n}$, and
$*\omega_{(n)}={\sqrt{-\det G}\over n!\,(10-n)!}\,
\varepsilon_{\nu_1\dots\nu_{10-n}\mu_1 \dots \mu_n} \,\omega^{\mu_1
\dots \mu_n}  dx^{\nu_1}\wedge\dots\wedge dx^{\nu_{10-n}}.$}:
\[
S_{\rm IIB} = \frac{1}{2 \kappa^2} \Bigg\{ \int d^{10} x~ \sqrt{-\det
 G}~ R - \frac{1}{2} \int \Big[ d \phi \wedge {}^* d \phi \,+\, {\rm
 e}^{- \phi} H_{3}  \wedge {}^* H_{3}\,+\, {\rm e}^{2 \phi}\, F_{1}
 \wedge {}^* F_{1}
\]
\begin{equation}
 + \,\,{\rm e}^{\phi} \,{\widetilde{F}}_{3} \wedge {}^*
 {\widetilde{F}}_{3} \,+\, \frac{1}{2}\, {\widetilde{F}}_{5} \wedge
 {}^* {\widetilde{F}}_{5}  \, -\,  C_{4} \wedge H_{3} \wedge F_{3}
 \Big] \Bigg\}
\label{tendim3}
\end{equation}
where  \be  H_{3} = d B_{2}~~~,~~~F_{1}=d C_{0}~~~,~~~ F_{3} = d
C_{2}~~~,~~~F_{5} = d C_{4}
\label{form2}
\ee  are, respectively, the field strengths of the NS-NS 2-form and
the 0-, 2- and 4-form potentials of the R-R sector, and: \be
{\widetilde{F}}_{3} = F_{3} + C_{0} \wedge H_{3}~~~~
,~~~~{\widetilde{F}}_{5} = F_{5} + C_{2} \wedge H_{3}~~.
\label{form3}
\ee       As usual, the self-duality constraint ${}^*
{\widetilde{F}}_{5}={\widetilde{F}}_{5}$  has to be implemented on
shell.

In order to find a classical solution corresponding to fractional
D3-branes, we have to add to the previous bulk action the
corresponding world-volume action, that we call generically $S_b$. By
varying the sum of the bulk and boundary action one can derive the
equations of motion for the various fields of type IIB
supergravity. One gets:
\begin{equation}
d {}^* d \phi + \frac{1}{2} {\rm e}^{-\phi} H_3 \wedge {}^* H_3 - {\rm
  e}^{2 \phi} F_1 \wedge {}^* F_1 - \frac{1}{2} {\tilde{F}}_3 \wedge
  {}^*  {\tilde{F}}_3 + 2 \kappa^2 \frac{\delta {\cal{L}}_b}{\delta
  \phi}=0
\label{dil89}
\end{equation} 
for the dilaton,
\begin{equation}
d \left( {\rm e}^{2 \phi} {}^* F_1 \right) - {\rm e}^{\phi} H_3 \wedge
{}^* {\tilde{F}}_3 + 2 \kappa^2 \frac{\delta{\cal{L}}_b}{\delta C_0}=0
\label{axi78}
\end{equation} 
for the axion,
\begin{equation}
d \left({\rm e}^{\phi} {}^* {\tilde{F}}_3  \right) + {\tilde{F}}_5
\wedge H_3 + 2 \kappa^2 \frac{\delta {\cal{L}}_b}{\delta C_2}=0
\label{c278}
\end{equation} 
for the R-R two-form,
\begin{equation}
d \left( {\rm e}^{-\phi} {}^* H_3 + {\rm e}^{\phi} C_0
  {\tilde{F}}_3\right) - {\tilde{F}}_5 \wedge F_3 + 2 \kappa^2
  \frac{\delta {\cal{L}}_b}{\delta B_2}=0
\label{b278}
\end{equation} 
for the NS-NS two-form field,
\begin{equation}
d {}^* {\tilde{F}}_5 + H_3 \wedge F_3 + 2 \kappa^2  \frac{\delta
{\cal{L}}_b}{\delta C_4}=0
\label{c478}
\end{equation} 
for the four-form R-R field and finally
\begin{equation}
R_{\mu \nu}- \frac{1}{4~\cdot
 4!}(\tilde{F}_5)_{\mu\rho\sigma\tau\delta}
 (\tilde{F}_5)_{\nu}^{\,\,\,\rho\sigma\tau\delta}+ 2 \kappa^2
 \frac{\delta \cal{L} }{ \delta G^{\mu\nu}}
 =\frac{1}{2}\left[\partial_{\mu}\phi\partial_{\nu}\phi +  {\rm
 e}^{2\phi} \partial_{\mu}C_{0}\partial_{\nu}C_{0}\right] + T_{\mu \nu}
\label{metric78}
\end{equation} 
for the metric, where:
\begin{equation}
T_{\mu \nu} =   \frac{{\rm e}^{-\phi}}{2\,\,3!} \left[ 3\,
H_{\mu\sigma\rho}H_{\nu}^{\,\,\,\sigma\rho} - \frac{G_{\mu\nu}}{4}
H^2\right]+ \frac{ {\rm e}^{\phi} } {2\,\,3!} \left[ 3 \,
(\tilde{F}_3)_{\mu\sigma\rho}(\tilde{F}_3)_{\nu}^{\,\,\,\sigma\rho} -
\frac{G_{\mu\nu}}{4} F_3^2\right] 
\label{tmunu78}
\end{equation} 
and ${\cal{L}}_b$ is the Lagrangian corresponding to the boundary
action. By using the quantities introduced in eq.(\ref{taug3}), it is
possible to rewrite the four equations for the dilaton, the axion and
the two 2-form potentials in terms of two complex equations as:
\begin{equation}
d {}^{*}d \tau + i {\rm e}^{\phi} d \tau \wedge {}^{*} d \tau +
\frac{i}{2} G_3 \wedge {}^{*}  G_3 = 2i \kappa^2 {\rm e}^{- \phi}
\left[ \frac{\delta {\cal{L}}_b}{\delta \phi} +i {\rm e}^{-\phi}
\frac{\delta {\cal{L}}_b}{\delta C_0} \right]
\label{dileq45}
\end{equation}
and
\begin{equation}
d {}^{*} G_3 + d \tau \wedge \left[i {\rm e}^{\phi} {}^{*} G_3 +
  {}^{*} H_3  \right] - i {\tilde{F}}_5 \wedge G_3 = - 2i \kappa^2
  \left[ \frac{\delta {\cal{L}}_b}{\delta B_2} - \tau \frac{\delta
  {\cal{L}}_b}{\delta C_2} \right]
\label{g3eq67}
\end{equation}
For a D3-brane we assume the following ansatz for the metric:
\begin{equation}
ds^{2}_{10} = H^{-1/2} \eta_{\alpha \beta}~ dx^{\alpha}~ dx^{\beta} +
H^{1/2}~ ds_{6}^{2}
\label{me429}
\end{equation}
and for the self-dual $5$-form field strength:
\begin{equation}
{\tilde{F}}_5 = d \left(\frac{1}{H} dx^0 \wedge \dots dx^3 \right) +
{}^* d \left(\frac{1}{H} dx^0 \wedge \dots dx^3  \right)
\label{f590}
\end{equation}
Notice that the six-dimensional space transverse to the D3-brane does
not need to be flat. The fact that the warp factors for the metric and
for the 5-form field strength are the same is a consequence of
supersymmetry, as it has been shown in the previous section. Using
eqs. (\ref{me429}) and (\ref{f590}), together with
eq.(\ref{identity}) and the following equation:
\begin{equation}
{}^* G_3 = - {}^{*6} G_3 \,\frac{1}{H}\,dx^0 \wedge \dots dx^3
\label{ide63}
\end{equation}
one can rewrite eq.(\ref{g3eq67}) as follows:
\begin{equation}
- d {}^{*6} G_3 \wedge \frac{1}{H} \,dx^0 \wedge \dots dx^3 = - 2i
\kappa^2 \left[ \frac{\delta {\cal{L}}_b}{\delta B_2} - \tau
\frac{\delta {\cal{L}}_b}{\delta C_2} \right]
\label{g3eq68}
\end{equation}
On the other hand, we can write:
\begin{equation}
G_3 = G_1 \wedge \omega_2~~,~~ G_1 \equiv d\gamma = dc + i db
\label{g359}
\end{equation}
where for a D3-brane we have taken $\tau$ to be constant and equal to
$i$ (notice that in terms of higher dimensional fields  we can write $\gamma =
\int_{{\cal C}_1} \left( C_2 + i  B_2\right)$). This is possible
because eq.(\ref{dileq45}) for $\tau$ is identically satisfied if we
assume eq.(\ref{identity}) as required by supersymmetry. On the other
hand since $\omega_2$  is a (1,1)-form, the condition that $G_3$
should be a $(2,1)$-form implies $\gamma$ to be an analytic function
of the complex variable $z_1 = x^4 + i x^5$(which, from now on, we
call $z$). By using the boundary  action in eq.(\ref{g3eq68}) one
arrives at the following equation for $\gamma$:
\begin{equation}
- d {}^{*6} d \gamma =  i \,\frac{2 \kappa^2 \tau_3}{2 \pi^2
\alpha'}\,  (N_1 - N_2) \,\delta^{(2)} (x) dx^4 \wedge dx^5 \wedge
^{*_4} \omega_2
\label{eqga87}
\end{equation}
where  the two-dimensional $\delta$-function is in the space spanned
by $x^4$ and $x^5$. From eq.(\ref{eqga87}) after some calculation one
gets:
\begin{equation}
\partial_i \partial^i \gamma = 2 \,\pi \,i \,K_3  \delta^{(2)} (x)
\label{eqga45}
\end{equation}
where $K_3$ is defined in eq.(\ref{bfi45}), $K_3 = 4 \pi g_s \alpha' (N_1 - N_2)$. 
The solution reads:
\begin{equation}
\gamma =\,i\, K_3 \log (z/z_{(1)})
\label{eqga46}
\end{equation} 
where $z_{(1)} \;=\; \epsilon \;{\rm e}^{- \pi\,/\,2 (N_1 - N_2)\,g_s}$ (this 
definition ensures that at $|z|=\epsilon$, which is a long-distance regulator 
for the $\log$, the field $\gamma$ has its correct background value).
Let us now consider the equation that determine the warp factor $H$.
Inserting the Ansatz (\ref{me429})-(\ref{f590}) into eq.(\ref{c478}),
we get:
\begin{equation}
\delta^{ij}\partial_i\partial_j H +
\frac{1}{2}\,|\partial_z\gamma|^2\,  \delta(x^6)\dots\delta(x^9) +
4\,\pi^3 Q_3 \,\delta(x^4)\dots \delta(x^9) = 0
\label{warp}
\end{equation}
where $Q_3$ is defined in eq.(\ref{qp876}), $Q_3 = 2 \pi g_s
(\alpha')^2 (N_1 + N_2)$. It is easy to verify that
eq.(\ref{metric78}) gives exactly the same  equation for the warp
factor $H$. Using standard technique it is possible  to integrate
eq.(\ref{warp}) obtaining:
\begin{equation}
H = 1+ \frac{Q_3}{r^4}+ \frac{K_3^2}{2\, r^4} \left[ \log
\left(\frac{r^4}{ \epsilon^2(r^2-|z|^2)}\right) - 1
+\frac{|z|^2}{r^2-|z|^2}\right]
\label{warp01}
\end{equation}
While the previous expressions, for $g_sN_1,g_sN_2 << 1$, reproduce
the large distance behaviour obtained from the boundary state in
section 4, they finally provide the complete supergravity solution we
were searching for. 

A closer look at the form of the warp factor $H$, shows that the
metric has a naked singularity at some point $r=r_0$ where indeed $H$
vanishes.   The singularity is of repulson type \cite{kal} because in
its vicinity the gravitational force, that is related to the gradient
of  $G_{00}$, is repulsive. The appearance of  these kind of
singularities is quite a general feature of supergravity solutions
corresponding to non-conformal sources and one expects that string
theory should be able  to resolve them. In this case, as we  discuss
in the next section, the singularity is resolved by an enhan\c{c}on
mechanism, similar to the one  originally discussed in
Ref.\cite{enhanc}, that excises the region close to the singularity
giving a regular solution in the region of the space time where it has
a physical meaning.

\section{The probe action and the $\mathbf{{\cal N}=2}$ gauge theory}
\label{8}
In this section we will try to see how much can supergravity tell us
about the gauge theories describing the low energy effective dynamics
of fractional branes. As we have discussed in section 3, fractional
D3-branes are described, in general, by non-conformal ${\cal N}=2$
super  Yang-Mills, at low energy. Therefore, answering the question
above could give non trivial insight on non-conformal extensions of
the gauge/gravity correspondence. While we will give a precise
meaning, at the gauge theory level, to all relevant physical
quantities entering the supergravity solution, we will also find out
that in order to get a prediction for the {\em full} moduli space of
the ${\cal N}=2$ gauge theory one should go beyond the pure
supergravity analysis. This is a quite general feature when
considering  non-conformal extensions of the gauge/gravity
correspondence,  and the answer, in this case, will be that
supergravity does indeed  encode the perturbative moduli space of the
gauge theory but it is not  able to include non-perturbative
corrections. As it will become clear in what  follows, the so-called
enhan\c{c}on \cite{enhanc} plays a crucial role  in all that. Indeed,
besides curing the naked singularity in a way that we are going to
discuss, it will also put a limit on the range  of validity of the
gauge/gravity correspondence pointing to a duality where  string states
play a role, even at low energies. After these anticipations, let us
now proceed to our analysis.

As explained in section 3, the low energy theory living on $N_1$
fractional D3-branes of type 1 and $N_2$ of type 2 is ${\cal N}=2$
super  Yang-Mills with gauge groups $SU(N_1) \times SU(N_2)$  and two
hypermultiplets transforming in the ($N_1,\bar N_2$) and ($\bar
N_1,N_2$), respectively. In order to get information on this gauge
theory from supergravity we shall use the probe technique. For a
review of this technique we refer to Ref.~\cite{joh}. 

Let us first consider a fractional D3-brane probe of type 1, carrying
a gauge field $F_{\alpha\beta}$ and slowly moving in the supergravity
background produced by $N_1$ fractional D3-branes of type 1 and $N_2$
of type 2. From the gauge theory point of view this corresponds to the
$SU(N_1)\times SU(N_2) \times U(1)$ broken phase of $SU(N_1+1)\times
SU(N_2)$  gauge theory and the probe gauge coupling  should equal the
gauge coupling of the first gauge group, $SU(N_1)$, at an energy scale
$\Lambda$ which is related to the distance $|z|$ at  which the probe
brane is taken far from the other branes. We  fix the static gauge and
study the world-volume action of the probe, regarding the transverse
coordinates as Higgs fields $\Phi^i= (2\pi\alpha')^{-1} x^i$, and
expanding up to quadratic terms in derivatives. By straightforward
computations we find that the probe action becomes:
\begin{equation}
S = S_0 + S_{\rm gauge}
\label{bound34}
\end{equation} 
where $S_0$ is just the same action of eq.(\ref{bou72}) with $p=3$
while       
\[
S_{\rm gauge} = -\frac{1}{4\pi g_s} \int d^4 x \sqrt{-\det
G_{\alpha\beta}}  ~\left\{\, \frac{1}{4} \,{\rm e}^{-\phi}\,G^{\alpha
\gamma} G^{\beta \delta} F_{\alpha \beta} F_{\gamma \delta} \,\,+
\right.  
\]
\begin{equation}
+\,\left.\frac{1}{2} \,G_{ij} G^{\alpha \beta} \partial_{\alpha}
\Phi^i  \partial_{\beta} \Phi^j \,\right\} \frac{1}{4\pi^2\alpha'}
\int_{{\cal C}_1} B_2 + \frac{1}{4\pi g_s}  \int d^4 x\,
\frac{1}{4}\,F_{\alpha \beta} {\widetilde{F}}^{\alpha \beta}
\,\frac{1}{4 \pi^2 \alpha'} \int_{{\cal C}_1} C_2
\label{boundg77}
\end{equation}
where $\widetilde{F}^{\alpha \beta}=(1/2)\,  \epsilon^{\alpha
\beta\gamma\delta} F_{\gamma\delta}$.  Inserting in $S_0$ the
supergravity  solution obtained in the section 7, one can easily see
that $S_0$ becomes independent of the distance between the probe and
the source branes that yield the classical solution. This is in
agreement with the fact that there is no interaction between the probe
and the source since fractional branes are BPS states and do not exert
any force on each other.

Considering now the above equation, we see that the dependence on the
function $H$ drops out in this case too, while the kinetic terms for
the gauge field strength $F_{\alpha\beta}$ and the  scalar fields
$\Phi^i$ have the same coefficient, in agreement with the fact that
the  gauge theory living on the brane has ${\cal{N}}=2$
supersymmetry. Indeed  one gets\footnote{For the sake of simplicity in
this formula (and subsequent ones) we define dimensionless 2-forms
$\hat B_2$ and $\hat C_2$ as  $\hat B_2 =
(2\pi\sqrt{\alpha'})^{-2}B_{2}$ and $\hat C_2 =
(2\pi\sqrt{\alpha'})^{-2}C_2$.}:
\begin{equation}
S_{gauge} = -\,\frac{1}{g_1(\mu)^2} \int d^4 x  \left\{ \frac{1}{2}
\partial_{\alpha} \Phi^i \partial^{\alpha} \Phi^i + \frac{1}{4}
F_{\alpha \beta} F^{\alpha \beta} \right\}  + \frac{\theta_1}{32
\pi^2}  \int d^4 x F_{\alpha \beta} {\tilde{F}}^{\alpha \beta}
\label{bound53}
\end{equation} 
where
\begin{eqnarray}
\frac{1}{g_1(\mu)^2 } &=& \frac{1}{4\pi g_s}\int_{{\cal C}_1} \hat B_2
= \frac{1}{g^2_{1}} + \frac{N_1-N_2}{4\pi^2} \log \mu
\label{runn23} \\
\nonumber \\ \theta_{1} &=& \frac{2\pi}{g_s}\int_{{\cal C}_1} \hat C_2
= 2 \,(N_1-N_2) \, \theta
\label{runn24}
\end{eqnarray}
are the effective Yang-Mills gauge coupling and the $\theta$-angle,
respectively. The renormalization group scale is defined by  $\mu =
|z|/\epsilon \equiv \Lambda/\Lambda_0$, while $g^{2}_{1}=8\pi g_s$ is
the bare coupling, i.e. the value of the gauge coupling at  the
ultraviolet cutoff $\Lambda_0 = (2\pi\alpha')^{-1} \epsilon$
\footnote{The  probe analysis automatically fixes the gravity/gauge
theory distance/energy  relation to be $|z| = 2\pi \alpha' \Lambda$
\cite{Peet:1998wn}.}. Eq.(\ref{runn23}) correctly predicts, from
supergravity, $g_{1}(\mu)$ to be the running coupling constant of an
${\cal{N}}=2$ supersymmetric gauge theory with gauge group $SU(N_1)$
and two hypermultiplets in the ($N_1,\bar N_2$) and ($\bar N_1,N_2$),
respectively ($N_2$ is a flavour index here).

By probing now the same background with a fractional D3-brane of type
2, one gets similar results. Eqs.(\ref{runn23}) and (\ref{runn24}), in
particular, become:
\begin{eqnarray}
\frac{1}{g_2(\mu)^2 } &=& \frac{1}{4\pi g_s}\,\left(1 - \int_{{\cal
C}_1}\hat B_2\right) = \frac{1}{g^2_{2}} - \frac{N_1-N_2}{4\pi^2} \log
\mu
\label{runn23b} \\
\nonumber \\ \theta_{2} &=& - \frac{2\pi}{g_s}\int_{{\cal C}_1} \hat
C_2 = 2 \,(N_2-N_1) \, \theta
\label{runn24b}
\end{eqnarray}
where $g_2(\mu)$ is the gauge coupling of the second gauge group
$SU(N_2)$ at the scale $\mu$ and $g^{2}_2 = 8\pi g_s$ (the role of
$N_1$ and $N_2$ is exchanged now, $N_2$ is a colour index while $N_1$
a flavour one). Again, the supergravity  prediction exactly agrees
with  the gauge theory expectations. Notice that the two
$\beta$-functions that one gets from eqs.(\ref{runn23}) and
(\ref{runn23b}), have opposite sign, according to the sign of
$N_1-N_2$. For $N_1>N_2$ (we will always use this convention in the
remaining part of this section), the first gauge theory is UV-free
while  the second one is IR-free. This follows from the relative
weights in the two theories of the matter present. In particular one
finds:
\begin{equation}
\beta(g_1) = - \frac{N_1 - N_2}{8 \pi^2} g_1(\mu)^3 \quad , \quad
\beta(g_2) = + \frac{N_1 - N_2}{8 \pi^2} g_2(\mu)^3
\end{equation}
Notice also that according to the general discussion in section 3, the
system we have probed actually corresponds to a bound state of say, $M
\equiv N_1-N_2$ fractional branes of type 1 and $N_2$ bulk
branes. This is the reason why the $\beta$-functions just depends on
$M$, the net number of fractional branes present. The gauge theory
living on a bulk brane  is conformal invariant and hence bulk branes
are expected not to give any contribution to the
$\beta$-function. This can also be seen by noticing that:
\begin{eqnarray}
\frac{1}{g_1(\mu)^2 } + \frac{1}{g_2(\mu)^2} &=& \frac{1}{4\pi g_s} \\
\frac{1}{g_1(\mu)^2 } - \frac{1}{g_2(\mu)^2} &=& \frac{1}{2 \pi g_s}
\left(\int_{{\cal C}_1} \hat B_2 - \frac{1}{2}\right)
\end{eqnarray}
where the sum of the couplings, corresponding to the contribution of
the  bulk branes, is not running, while the difference, measuring the
(net amount of) fractional branes contribution, is running according
to the (fluctuation of the) $B_2$-flux.

The previous results show that supergravity provides the perturbative
moduli space of the gauge theory (which is exact at one loop, in this
case). But what about non-perturbative contributions?  Here is where
the enhan\c{c}on comes into play. We have seen that, because of the
presence of a non-vanishing background $B_{2}$-flux \cite{bflux}, the
fractional branes are in general tensionful. On the other hand, since
the factor in front of the gauge kinetic term  in eq.(\ref{bound53})
is the same that gives the effective tension of the brane, as one can
see comparing eqs. (\ref{boundg77}) and (\ref{bou72}), the tension of
a fractional brane is running precisely as the gauge coupling constant
in eq.(\ref{runn23}). From it one sees that on the geometric locus
defined by
\begin{equation}
z_{(1)} \;=\; \epsilon \;{\rm e}^{- \pi\,/\,2 (N_1 - N_2)\,g_s}
\label{enh}
\end{equation}
the type 1 brane probe becomes tensionless! This locus is known as the
enhan\c{c}on \cite{enhanc}. This is the point where the fluctuation of
the $B_2$-field cancels precisely its background value.  The vanishing
of the probe tension indicates  that at the distance $z_{(1)}$  new
light (string) degrees of freedom come into
play~\cite{Polchinski:2000mx},  meaning that the supergravity
approximation leading to the solution described in the previous
section is not valid in the region of space-time $\rho < z_{(1)}$. So,
the solution we have found  makes sense only at distances bigger than
the enhan\c{c}on and the unwanted repulson singularity, which is
cloaked inside it (one can show this is the case for any choice of the
parameters), is then excised.

One can immediately recognize what is the meaning of the enhan\c{c}on
from the gauge theory point of view.  In fact eq.(\ref{runn23}) shows
that the enhan\c{c}on is the scale $\Lambda_{(1)} = (2\pi\alpha')^{-1}
z_{(1)}$ where the  gauge coupling $g_1(\mu)$ diverges and where
non-perturbative corrections become relevant ($\Lambda_{(1)}$ then
corresponds to the dynamically generated scale). This automatically
implies that the supergravity solution is only able to reproduce the
perturbative moduli space of the gauge theory, while the appearance of
the enhan\c{c}on prevents from using the classical solution to analyze
the strong-coupling properties of the gauge theory, where instanton
effects should become relevant \cite{SW}.

A similar reasoning can be repeated for the probe brane of type 2. The
only subtle point one should bare in mind is that the $\beta$-function
is now IR-free. The consequence is that the corresponding enhan\c{c}on
appears in the UV, which indeed now corresponds to the  strongly
coupled region of the theory. This region is, however, not really
important in our present analysis because, for any value of the
parameters,  it is always bigger then the UV cut-off $\Lambda_0 =
(2\pi\alpha')^{-1}\epsilon$ and  hence out of reach of the
supergravity solution, whose logarithmic running for twisted fields is
regulated at $|z|=\epsilon$. Indeed, the expression of the type 2
enhan\c{c}on is
\begin{equation}
z_{(2)} \;=\; \epsilon \;{\rm e}^{\pi\,/\,2 (N_1 - N_2)\,g_s} >
\epsilon
\label{enh2}
\end{equation}
Summarizing, the physical picture one ends up with is that
supergravity reproduces the gauge theory of $SU(N_1)\times SU(N_2)$
between the UV-cutoff $\Lambda_0$ and the type 1 dynamically generated
scale  $\Lambda_{(1)}$ (the enhan\c{c}on), while the extreme  IR and
UV regions are not accessible by supergravity.

From the previous considerations, one should suspect the existence of
a precise relation between the twisted field $\gamma$ and the period
matrix $\tau$ of Seiberg-Witten \cite{SW}. To find it out we should
consider, for  each gauge group, the explicit expression of the
prepotential ${\cal F}$,  compute it in the corner of the moduli space
consistent with our probe  analysis, and finally recall the relation
between ${\cal F}$ and $\tau$, namely $\tau_{lm}=\partial^2{\cal
F}/\partial a_l\,\partial a_m$, where $a_l$ and $a_m$ are the moduli
of the gauge theory ($l,m=1,...,N_I+1$ and $I=1,2$  since we have two
gauge groups). For any of the two gauge groups, the  corresponding
perturbative prepotential reads (see for instance Ref.\cite{dp}):
\begin{eqnarray}
\label{Fgen}
{\cal F}_{pert} &=& \frac{i}{8\pi} \sum_{l,m=1}^{N_I+1} \left(a_l -
a_m\right)^2 \log\frac{(a_l-a_m)^2}{\Lambda_{(I)}^2}\nonumber\\
& -& \frac{i}{8\pi}
\sum_{l=1}^{N_I+1} \,\sum_{k=1}^{2 N_J} (a_l + M_k)^2
\log\frac{(a_l+M_k)^2}{\Lambda_{(I)}^2}
\end{eqnarray}
where $\Lambda_{(I)}$ is the dynamically generated scale (see above),
and $M_k$ are the masses of the hypermultiplets corresponding to
strings stretched between branes of different types (the sum over $k$
goes up to the number of hypermultiplets which is 2 times the number
of flavours which for the gauge group $U(N_I)$ is indeed $N_J$). 

The type $I$ probe analysis corresponds to the breaking $SU(N_I+1)
\rightarrow SU(N_I) \times U(1)$ of  the corresponding gauge group,
which, in terms of the moduli $a_l$ means that one modulus, say
$a_{N_I+1}\equiv a$, is taken to be much bigger than the others. This
simplifies the prepotential in eq.(\ref{Fgen}) as:
\begin{equation}
\label{Fprobe}
{\cal F} = \frac{i}{4\pi} \left(N_I-N_J\right) a^2
\log\frac{a^2}{\Lambda_{(I)}^2}
\end{equation}
which implies that:
\begin{eqnarray}
\label{tau0}
&&\tau_{l,m} \sim 0 \qquad \mbox{with} \quad l,m = 1,...,N_I \\
\label{tau1}
&&\tau_{N_I+1,N_I+1} \equiv \tau_I = \frac{4\pi}{g_I(\mu)^2} \,i +
\frac{\theta_I}{2\pi} = \frac{i}{\pi} (N_I - N_J)  \log
a/\Lambda_{(I)}
\end{eqnarray}
From the above equation one gets the precise relation between the
complex twisted field  $\hat \gamma(z)$ and the above computed
$\tau_I$ to be just:
\begin{equation}
\label{tg}
\hat \gamma_I(z) = g_s \tau_I(z)
\end{equation}
where we have defined 
\begin{equation}
\hat \gamma_1 = \int_{{\cal C}_1} \left(\hat C_2 + i \hat
B_2\right)~~~,~~~\hat \gamma_2 = \int_{{\cal C}_0} \left(\hat C_2 + i
\hat B_2 + 2\pi i \alpha' {\cal F}\right)
\label{bbb98}
\end{equation}
Again, as already stressed, the above identification holds from the
supergravity solution only up to the perturbative part of $\tau$.

For the sake of clarity, we finally summarize  in table \ref{gagrc}
all the relations between supergravity and gauge theory quantities.
\begin{table} [ht] 
\vskip 10pt
\begin{center}
\begin{tabular}{|c|c|c|}
\hline GRAVITY       &          &  GAUGE THEORY \\    \hline
Transverse coordinates   &  $x^i \longleftrightarrow \Phi^i$  & Higgs
field \\ \hline $B_2$-flux through ${\cal C}_1$ & $\int_{{\cal C}_1}
B_2 \longleftrightarrow g_1(\mu)$ & $U(N_1)$ gauge coupling \\  \hline
$B_2$-flux through ${\cal C}_0$ & $\int_{{\cal C}_0}
B_2\longleftrightarrow g_2(\mu) $ & $U(N_2)$ gauge coupling \\  \hline
$C_2$-flux through ${\cal C}_1$ & $\int_{{\cal C}_1} C_2
\longleftrightarrow \theta_1$ & $U(N_1)$ $\theta$-angle  \\    \hline
$C_2$-flux through ${\cal C}_0$ & $\int_{{\cal C}_0} C_2
\longleftrightarrow \theta_2$ & $U(N_2)$ $\theta$-angle  \\    \hline
IR-regulator  & $\epsilon \longleftrightarrow \Lambda_0$  & UV-cutoff
\\    \hline Enhan\c{c}on  & $ z_{(1)} \longleftrightarrow \Lambda_{(1)}$
& Dynamically generated scale  \\    \hline
\end{tabular}
\end{center}
\caption{\small Correspondence between gravity and gauge theory
parameters.  The precise numerical relations can be found in  the main
text.}
\label{gagrc}
\end{table}

Let us end this section with few final observations. As it has been
discussed in Ref.\cite{marco}, by computing the flux of the untwisted
field strength $\tilde F_5$ through a surface which intersects the
$z$-plane on some given curve $\Sigma$, one gets:     \be     \Phi
(\tilde F_5) = 2 \pi^2 g_s \,\left( N_1 + N_2  + \frac{g_s}{2\,\pi}\,
(N_1 -N_2)^2 \log \frac{|z|}{\epsilon}\right)    \ee  From the above
equation one can see that the 5-form flux is running. This is a
general feature of supergravity solutions generated by non-conformal
sources (see for instance Refs.\cite{KLEBA2,KLEBA3,HKO}), indicating
that the effective untwisted  charge is decreasing through the IR, the
decreasing being proportional to the net amount of fractional branes
present, $M=N_1-N_2$\footnote{Notice that at the enhan{\c{c}}on the
fractional brane contribution  vanishes and the flux becomes
proportional to the amount of bulk charge  contribution,
i.e. $\Phi(\tilde F_5) = 4 \pi^2 g_s N_2$ (recall that we are choosing
$N_1 > N_2$ and therefore $N_2$, according to the discussion in
section 3, is the actual number of bulk branes present in the bound
state).}. Qualitatively this is a correct result since the untwisted
charge corresponds to the  number of degrees of freedom of the dual
gauge theory, and these are expected to diminish through the IR. It is
also correct that this decreasing is proportional to $M$, since
fractional branes make the  gauge theory non-conformal and hence are
the ones responsible for the running.  However, in order to make a
more quantitative matching, one should have a better understanding of
the physics of the enhan\c{c}on. There have been various efforts in
trying to understand at a deeper level the role of the enhan\c{c}on
for this and other systems (see in particular
Refs.\cite{Polchinski:2000mx,Aharony,Petrini:2001fk} for interesting
discussions on this point). However, quantitative results to go beyond
the above successful perturbative analysis, have not yet been
obtained.  The fact that fractional D-brane probes vanish at the
enhan\c{c}on, has suggested  the idea \cite{enhanc} that it is not
possible to build up a source made of  fractional D-branes located at
the origin $r=0$. Rather, the constituent branes  are smeared on the
enhan\c{c}on shell and branes that are coupled to the twisted fields
and that therefore become tensionless at the enhan{\c{c}}on cannot
enter inside the enhan\c{c}on region. In this way it is clear that,
while  the exterior solution, due to Gauss' theorem, is of course
unchanged, the interior one could look completely different. This
picture has been shown to  agree both with excision criteria
\cite{exc,Merlatti:2001gd} and, more concretely, with some solid gauge
theory consistency checks which have been done using the SW curve
\cite{Petrini:2001fk}.  At the same time the enhan\c{c}on cannot be
the end of the story. Indeed,  at the enhan\c{c}on the moduli space
metric, which is proportional to $g_1(\mu)^{-2}$, vanishes. According
to Seiberg-Witten \cite{SW}, this cannot be the case in the full
quantum moduli space of ${\cal N}=2$ super Yang-Mills, and in fact you
should take into account the instanton corrections, which become
relevant at strong coupling, and that give rise to a positive definite
moduli space metric. This is consistent with the previous supergravity
analysis, which indicates the presence of new light degrees of freedom
at the enhan\c{c}on scale. By including them in the low energy
effective action, one should get back an enhan\c{c}on free and
singularity free  solution, as discussed recently in Ref.\cite{ben}
and, on the side of the gauge theory, one should recover the
non-perturbative corrections. 

Let us end with the following (qualitative) observation.  From the
gauge theory side, the effective gauge theory will receive corrections
proportional to powers of the one-instanton contribution to the
partition function
\begin{equation}
\mbox{exp}\left( 2\pi i \tau_1 \right) =
\left(\frac{\Lambda_{(1)}}{z}\right)^{2 (N_1 - N_2)}
\end{equation}
As noticed in Ref.\cite{Klebanov:2000rd}, on the string theory side
such effects can be due to fractional D-instantons (which are D1
Euclidean branes wrapped on the vanishing cycle ${\cal C}_1$ and which
indeed become tensionless at the enhan\c{c}on), whose action is indeed
\begin{equation}
\mbox{exp}\left( 2 \pi i \,\frac{\hat \gamma_1}{g_s}\right) \equiv
\mbox{exp}\left( 2\pi i \tau_1 \right)
\end{equation}
Unfortunately, it has not been possible until now to make this
argument quantitative and in particular to determine the coefficients
of the instanton corrections~\cite{fu}.

Again, all what we have been discussing in this section can be
extended to  more general orbifolds of the complete ADE series. We
refer again to Ref. \cite{marco} for a complete treatment of these
more general cases.

\section*{Acknowledgments}
\addcontentsline{toc}{section}{\numberline{}Acknowledgments}  
First of all we would like to thank the co-authors of
Refs.\cite{Bertolini:2000dk,d3d7} with whom we share  the results
presented in this paper, for the friendly and productive
collaboration.  We thank also  M. Bill\`o, M. Bonini, G. Bonelli,
G. Ferretti,  P. Fr\`e, T. Harmark, A. Liccardo, N. Obers, F. Pezzella, 
R. Russo, M. Trigiante and  A. Westerberg for many useful discussions. 
This work is partially
supported by the EC RTN contract HPRN-CT-2000-00131. M.B. is supported
by a EC Marie Curie Postdoc Fellowship under contract number
HPMF-CT-2000-00847.

\addcontentsline{toc}{section}{\numberline{}References}

\end{document}